\documentclass{aa}
\usepackage{graphics}
\usepackage{latexsym}
\begin{document}

   \thesaurus{03     
	      (11.17.3;  
               13.09.1;  
               13.18.1)} 

 \title{The Far-Infrared emission of Radio Loud and Radio Quiet Quasars.}

   \subtitle{}

   \author{M. Polletta\inst{1,2},
         T.J.-L. Courvoisier\inst{1,2},
         E.J. Hooper\inst{3},
          \and
         B.J. Wilkes\inst{3}
          }

   \offprints{Maria.Polletta@obs.unige.ch}

   \institute{Geneva Observatory, Ch. des Maillettes 11, CH-1290 Sauverny, Switzerland
         \and
             Integral Science Data Centre,  Ch. d'Ecogia 16, CH-1290 Versoix, Switzerland
         \and
             Harvard-Smithsonian Center for Astrophysics, Cambridge, MA, USA
             }

   \date{Received ...... .. ....; accepted ...... .. ....}

   \titlerunning{Far-IR of Radio Loud and Radio Quiet Quasars}
   \authorrunning{Polletta M. et~al.}
   \maketitle

   \begin{abstract}

Continuum observations at radio, millimetre, infrared and soft X-ray
energies are presented for a sample of 22 quasars, consisting of flat and
steep spectrum radio loud, radio intermediate and radio quiet objects. The
primary observational distinctions, among the different kinds of quasars in
the radio and IR energy domains are studied using large observational
datasets provided by ISOPHOT on board the Infrared Space Observatory, by
the IRAM interferometer, by the sub-millimetre array SCUBA on JCMT, and
by the European Southern Observatory (ESO) facilities IRAC1 on the 2.2 m
telescope and SEST.

The spectral energy distributions of all quasars from radio to IR energies
are analyzed and modeled with non-thermal and thermal spectral components.

The dominant mechanism emitting in the far/mid-IR is thermal dust emission
in all quasars, with the exception of flat spectrum radio loud quasars for
which the presence of thermal IR emission remains rather uncertain, since it
is difficult to separate it from the bright non-thermal component. The dust
is predominantly heated by the optical/ultraviolet radiation emitted from
the external components of the AGN. A starburst contributes to the IR
emission at different levels, but always less than the AGN ($\leq$ 27\%).
The distribution of temperatures, sizes, masses, and luminosities of the
emitting dust are independent of the quasar type.


   \end{abstract}

%

\section{Introduction}
\label{intro}

Radio quiet and radio loud (not blazar) quasars (RQQ and RLQ, respectively)
have similar spectral properties in the ultraviolet (UV), optical, and
infrared (IR), but their radio powers differ by several orders of magnitude
(\cite{Elvis94}). This divergence takes place at millimetre (mm) wavelengths. At
these wavelengths the contribution from two emission components merge,
namely the synchrotron emission dominant in the radio domain and thermal
emission from cool dust (30-50 K) in the far-IR (\cite{BA89}). It is
still not entirely clear whether the distinction between RLQ and RQQ is a
consequence of differences in their central engines or whether it merely
reflects differences in their environments. The primary observational
distinctions in the IR domain, and the proposed physical mechanisms to
explain them are studied here, using the new insights provided by 
Infrared Space Observatory\footnote{ISO is an ESA project with instruments
funded by ESA Member States (especially the PI countries: France, Germany,
the Netherlands and the United Kingdom) and with the participation of ISAS
and NASA.} (ISO;
\cite{Kessler96}) measurements.

\subsection{The Radio emission}


Two main types of RLQ can be distinguished on the basis of their radio
spectrum: the flat spectrum radio loud quasars (FSRQ), and the steep
spectrum radio loud quasars (SSRQ). FSRQ show highly-collimated
structures and very compact features, with flat or inverted radio spectra.
SSRQ have radio spectra dominated by synchrotron emission from extended
radio lobes. The lobes and a radio core in the centre of these objects
are signs of a relativistic jet. According to the unified scheme
(\cite{Barthel89};~\cite{Urry95}) FSRQ are the counterparts of SSRQ in which
the jet is aimed at the observer.

The origin of the much weaker radio emission in RQQ is far less certain. The
majority of the total radio emission from the RQQ comes from the compact
features in the nucleus ($<$ 1 kpc for unresolved regions, and at least 2
kpc for the resolved ones) rather than the body of the host galaxy
(\cite{Kukula98}). It has been proposed that the activity in RQQ is supplied
by a starburst, i.e. thermal bremsstrahlung and synchrotron emission coming
from strongly radiative supernov\ae\, and supernov\ae\, remnants (SNRs) in a
very dense environment where shocks accelerate electrons
(\cite{Terlevich92}). Alternatively, if the energy supply arises from
accretion onto a massive black hole, the radio emission from RQQ (as in RLQ)
is caused by radio jets, but the bulk kinetic power of these jets is for
some reason $\sim$ 10$^3$ times lower than those of RLQ (\cite{Miller93}).
This second hypothesis seems to be favored by recent studies, because
of high brightness temperatures calculated (typical SN/SNRs have
T$_\mathrm{B}\leq10^{5}$K), by the evidence of a pc-scale jet
(\cite{Blundell98}) and by observations of flat/inverted and variable radio
spectra (\cite{BLA96}).

Recently, quasars with intermediate radio luminosities have been discovered
and labeled Radio Intermediate Quasars (RIQ)
(\cite{Francis93},~\cite{Falcke95}). RIQ may represent the Doppler boosted
counterparts of radio quiet quasars. This hypothesis is suggested by the
variability observed at radio wavelengths (\cite{Falcke96}).

\subsection{The Infrared emission}

The presence of a dominant thermal (circumnuclear dust emission), or 
non-thermal (synchrotron radiation from the AGN) component in the IR
continuum of quasars is still debated.

Many attempts to establish the origin of the IR emission in RLQ and RQQ have
been performed through observations in the sub-millimetre (sub-mm) of
quasars detected by IRAS (RLQ in \cite{Chini89a}, and \cite{ABA90}; RQQ
in \cite{CKB89b}, \cite{Barvainis92}, \cite{Hughes93}, and \cite{Hughes97};
and both in \cite{Andreani99}, this last work is the only one based on an
optically selected sample). The main test applied to recognize the
presence of thermal emission in the IR spectra of the objects was based on
the slope of the continuum emission (F$_{\nu}\propto\nu^{\alpha}$)
connecting the far-IR and sub-mm data. A steep, $\alpha\,>$ 2.5, continuum
is strong evidence for thermal dust emission. Most of the sources studied
have $\alpha\,<$ 2.5 and are consistent with a dominant self-absorbed
synchrotron emission component. However, some RQQ have spectral slopes as
steep as $\alpha$ = 4.35, which, along with observations of strong molecular
gas (CO) emission (\cite{Barvainis97}), give strong support to a thermal
mechanism as the origin of the far-IR component in RQQ.

Among the RLQ, $\alpha$ is, at most, $\simeq$ 0.9 for the FSRQ, and 1.1 for
the SSRQ (\cite{Chini89a}). Variability, shape of the continuum
spectral energy distribution and, in some cases polarization, indicate that
the radio, mm and far-IR emission of FSRQ is dominated by the synchrotron
process (\cite{Lawrence91}). On the contrary, many SSRQ show evidence of
thermal emission: their far-IR spectra are brighter than extensions of the
radio emission (\cite{ABA90}), suggesting a different origin than the
non-thermal radio component; and the flux is constant, consistent with it
arising from a region much larger than a light year (\cite{Edelson87}).
Moreover, the spectral energy distributions of some RLQ, both FSRQ and SSRQ,
show evidence for a galaxy component: reddening, residual starlight,
molecular gas (\cite{Scoville93}), and some thermal dust emission in the
near-IR (\cite{Barvainis87}). Both components, non-thermal synchrotron
radiation and thermal dust emission, are probably present at IR wavelengths,
as observed in the RLQ 3C273 (\cite{Robson86};~\cite{Barvainis87}).

\subsection{Relation between the Radio and Infrared emission}

A tight, linear correlation is observed between the far-IR flux and the
radio fluxes in AGN (\cite{Sopp91}), suggesting a common origin. RQQ and RLQ
occupy well defined regions in Log($L$(IR))--Log($L$(Radio)) space, and show
a relation with a similar slope, just shifted to higher radio power by a
factor $\sim$10$^{4}$ in RLQ. RQQ show a similar relation as spirals,
starbursts, and ultra luminous IR galaxies (ULIRG), suggesting that their IR
emission may arise from sufficiently energetic star formation in the host
galaxy (\cite{Sopp91}). However, the majority of the bolometric luminosity
in over half of known ULIRG seems to arise from a buried AGN
(\cite{Sanders99}). Additionally, the variable and flat spectra, and high
brightness temperatures shown by many RQQ at radio frequencies
(\cite{BLA96}) suggest that the radio emission is related to the AGN rather
than to a starburst.

\subsection{Proposed scenarios}

The unified model (\cite{Barthel89}; \cite{Urry95}) predicts that
similar disk--like dust distributions exist in both RQQ and RLQ.
Orientation of the active nucleus, environment, and jet luminosity all
affect the relative contributions of thermal and non-thermal sources to
the observed infrared luminosity (\cite{Chini89a}). 

Other scenarios have been proposed to explain the large differences in radio
power between RQQ and RLQ: different spin of the central black hole
(\cite{Wilson95}), or different morphological type of the host galaxy.
Indeed, different radio powers are expected if one population of objects is
fueled by mergers (ellipticals) and one is fueled by mostly internal
processes within the galaxy (spirals) (\cite{Wilson95}). However, recent
studies on the host galaxies of quasars indicate that the host galaxies of
RQQ are in several cases elliptical and not always spiral galaxies
(\cite{Taylor96}). 

\subsection{Open issues}

A better knowledge of the radio and IR properties of quasars is required to
test the unified model predictions, and answer the following questions:
\begin{enumerate}
\item What is the dominant mechanism emitting at IR wavelengths
      in RLQ and RQQ?
\item Do RLQ and RQQ have the same dust properties (temperature, source
      size, mass, and luminosity)?
\item Does an interplay between the radio and the IR components exist?
\end{enumerate}

These questions can be addressed through the study of the spectral energy
distributions (SED) of RLQ and RQQ. Here, we present the SEDs from radio to
IR frequencies of a sample of 22 AGN (7 RQQ, 11 RLQ, 2 radio galaxies (RG)
and 2 RIQ). The selected sample, even if incomplete and heterogeneous, is
useful to address these questions thanks to several properties
characterizing the sample (steep/flat radio spectra, radio
loudness/quietness), and to the large amount of photometric data available
in the radio, mm/sub-mm and IR domains. This work is based mainly on IR data
provided by ISO. ISO data reduce the frequency gap between sub-mm and far-IR
observations, better sample the IR spectral band with a larger number of
filters than previous instruments, and increase the number of detected
objects thanks to a higher sensitivity. The study of the IR emission of
quasars will be extended in the future with the results of the European and
of the U.S. ISO Key Quasar Programs providing a similar coverage of the IR
SED for a larger sample of quasars (see first results in Haas et~al. (1998),
and Wilkes et~al. (1999)).

\section{Observational dataset}
\label{dataset}

Source names, coordinates, and redshifts of the selected sample were 
taken from the NASA Extragalactic Database (NED)\footnote{The NASA/IPAC
Extragalactic Database (NED) is operated by the Jet Propulsion Laboratory,
California Institute of Technology, under contract with the National
Aeronautics and Space Administration.}, and are listed in Table~\ref{BasProp}.
Infrared observations were obtained for 18 of the sources with the Imaging
Photopolarimeter on ISO (ISOPHOT;~\cite{Lemke96}), and 3 were observed with
IRAC1 on the 2.2 m ESO/MPE telescope.  Millimetre and sub-mm data were
obtained for 10 objects with the IRAM interferometer at Plateau de Bure in
France (\cite{Guilloteau92}), the Sub-millimetre Common User Bolometer Array
(SCUBA;~\cite{Holland99}) on the James Clerk Maxwell Telescope\footnote{The
James Clerk Maxwell Telescope is operated by the Joint Astronomy Centre on
behalf of the Particle Physics and Astronomy Research Council of the United
Kingdom, the Netherlands Organization for Scientific Research, and the
National Research Council of Canada.} and the Swedish ESO Sub-mm Telescope
(SEST) of the European Southern Observatory (ESO) at La Silla. The
instruments used for each object are indicated in a footnote to
Table~\ref{BasProp}.

   \begin{table}
      \caption[]{Basic properties of the sample}
         \label{BasProp}   
\begin{tabular}{@{}l@{}c@{}c@{}l@{}c@{}}
\hline
\noalign{\smallskip}
\multicolumn{1}{c}{Source} &
\multicolumn{1}{c}{RA (J2000)} &
\multicolumn{1}{c}{Dec (J2000)} &
\multicolumn{1}{l}{Type$^{\dagger}$} &
\multicolumn{1}{c}{z} \\
\multicolumn{1}{c}{Name} &
\multicolumn{1}{c}{$h\;\;m\;\;s$} &
\multicolumn{1}{c}{$\degr\;\;\arcmin\;\;\arcsec$} &
\multicolumn{1}{l}{}&
\multicolumn{1}{c}{}\\
\noalign{\smallskip}
\hline
\noalign{\smallskip}
\object{3C 47}$^{1,4}$        &  01 36 24 &   +20 57 27  & SSRQ  &  0.425 \\  
\object{PKS 0135$-$247}$^{1}$ &  01 37 38 & $-$24 30 53  & FSRQ  &  0.831 \\  
\object{PKS 0408$-$65}$^{1,2}$&  04 08 20 & $-$65 45 09  & RG    &  ...   \\  
\object{PKS 0637$-$75}$^{1,2}$&  06 35 47 & $-$75 16 17  & FSRQ  &  0.653 \\  
\object{PG 1004+130}$^{2}$    &  10 07 26 &   +12 48 56  & SSRQ  &  0.240 \\  
\object{PG 1048$-$090}$^{2}$  &  10 51 30 & $-$09 18 10  & SSRQ  &  0.344 \\  
\object{4C 61.20}$^{1}$       &  10 52 33 &   +61 35 20  & SSRQ  &  0.422 \\  
\object{PG 1100+772}$^{1,5}$  &  11 04 14 &   +76 58 58  & SSRQ  &  0.312 \\  
\object{PG 1103$-$006}$^{2}$  &  11 06 32 & $-$00 52 52  & SSRQ  &  0.423 \\  
\object{PG 1216+069}$^{1,2}$  &  12 19 21 &   +06 38 38  & RIQ$^{a}$   &  0.331 \\  
\object{PG 1352+183}$^{1}$    &  13 54 36 &   +18 05 18  & RQQ   &  0.152 \\  
\object{PG 1354+213}$^{1}$    &  13 56 33 &   +21 03 51  & RQQ   &  0.300 \\  
\object{PG 1435$-$067}$^{1,3}$&  14 38 16 & $-$06 58 21  & RQQ   &  0.129 \\  
\object{PG 1519+226}$^{1}$    &  15 21 14 &   +22 27 43  & RQQ   &  0.137 \\  
\object{PG 1543+489}$^{1,5}$  &  15 45 30 &   +48 46 09  & RQQ   &  0.400 \\  
\object{PG 1718+481}$^{1}$    &  17 19 38 &   +48 04 13  & RIQ$^{a}$   &  1.084 \\  
\object{B2 1721+34}$^{1}$     &  17 23 20 &   +34 17 58  & SSRQ  &  0.206 \\  
\object{HS 1946+7658}$^{1,4}$ &  19 44 55 &   +77 05 52  & RQQ   &  3.020 \\  
\object{3C 405}$^{1}$         &  19 59 28 &   +40 44 01  & NLRG  &  0.056 \\  
\object{B2 2201+31A}$^{1}$    &  22 03 15 &   +31 45 38  & FSRQ  &  0.295 \\  
\object{PG 2214+139}$^{3}$    &  22 17 12 &   +14 14 21  & RQQ   &  0.066 \\  
\object{PG 2308+098}$^{1,3}$  &  23 11 18 &   +10 08 15  & SSRQ  &  0.433 \\  
\noalign{\smallskip}                                                    
\hline         
\end{tabular}\\                                                           
$^{\dagger}$ SSRQ: Steep Spectrum Radio Loud Quasar; FSRQ: Flat Spectrum
Radio Loud Quasar; RQQ: Radio Loud Quasar; RIQ: Radio Intermediate Quasar;
NLRG: Narrow Line Radio Galaxy; RG: Radio Galaxy.

$^{1}$ Observed with ISOPHOT; $^{2}$ Observed with SEST; $^{3}$ Observed with
IRAC1; $^{4}$ Observed with IRAM; $^{5}$ Observed with SCUBA. 

$^{a}$ Classification in Falcke et~al. (1996).
\end{table}

The observational details (observing date, wavelength, and measured flux
density) of IRAC1, SEST, IRAM, and SCUBA observations are reported in
Table~\ref{OtherInstRes}. The data obtained with the SEST telescope were
reduced to outside the atmosphere, corrected for the gain elevation
characteristic of the telescope, and calibrated with Uranus. All IRAM
observations were performed in compact configuration. All five antenna were
used during most of the observations, with the exceptions indicated in a
footnote of Table~\ref{OtherInstRes}. The calibrators were 3C 454.3 for 3C
47, and 22040+420 and 1928+738 during the two observations of HS 1946+7658.
SCUBA observations yielded good results at 850 $\mu$m only due to
marginal weather. The fluxes were calibrated using the canonical gain value
of 220 Jy/V, since a calibration scan at 850$\mu$m was not done. The chosen
value is relatively insensitive to the weather and should be good to within
20\%. ISOPHOT observations and data reduction are described in the next
section. These new observations were supplemented with literature data from
radio to near-IR. We have also collected data at soft X-ray energies. For
reasons of homogeneity we collected only ROSAT data, available for most of
the sources. From published soft X-ray spectra we derived the flux at 1 keV
corrected for absorption. The observed absorption is always compatible with
the galactic absorption. The list of references from which data were
retrieved, for each object, is reported in Table~\ref{list_of_ref}.

   \begin{table}
      \caption[]{IRAC1, SEST, IRAM and SCUBA results}
         \label{OtherInstRes}
\begin{tabular}{@{}l@{}l@{}l@{}r@{}r@{}}
\hline
\noalign{\smallskip}
\multicolumn{1}{c}{Source Name} &
\multicolumn{1}{c}{Instrum.} &
\multicolumn{1}{l}{Obs. Date} &
\multicolumn{1}{c}{$\lambda$} &
\multicolumn{1}{c}{$\;\;\;\;\;\;\;$F$_{\nu}^{\dagger}\;\;\;\;$} \\
\multicolumn{1}{c}{} &
\multicolumn{1}{c}{} &
\multicolumn{1}{l}{($d\;m\;y$)} &
\multicolumn{1}{c}{($\mu$m)} &
\multicolumn{1}{c}{(mJy)} \\
\noalign{\smallskip}
\hline
\noalign{\smallskip}
3C 47            & IRAM  & 14 07 98$^{1,2}$ & 1300 & 17.0$\pm$2.3 \\
                 &       & 14 07 98$^{1,3}$ & 3000 & 30.8$\pm$0.6 \\
\noalign{\smallskip}
PKS 0408$-$65    & SEST  & 30 11 95        & 1300 & $<$30        \\
\noalign{\smallskip}
PKS 0637$-$752   & SEST  & 30 11 95        & 1300 & 724$\pm$37   \\
                 &       &                 & 1300 & 905$\pm$31   \\
\noalign{\smallskip}
PG 1004+130      & SEST  & 30 11 95        & 1300 & $<$39        \\
\noalign{\smallskip}
PG 1048$-$090    & SEST  & 30 11 95        & 1300 & $<$18        \\
\noalign{\smallskip}
PG 1100+772      & SCUBA & 17 01 99        &  850 & 6.8$\pm$2.4  \\
\noalign{\smallskip}
PG 1103$-$006    & SEST  & 30 11 95        & 1300 & $<$27        \\
\noalign{\smallskip}
PG 1216+069      & SEST  & 30 11 95        & 1300 & $<$45        \\
\noalign{\smallskip}
PG 1435$-$067    & IRAC1 & 21 06 96        &  3.7 & 12.1$\pm$0.4 \\
                 &       &                 &  3.7 & 12.2$\pm$0.5 \\
\noalign{\smallskip}
PG 1543+489      & SCUBA & 17 01 99        &  850 & $<$5.4       \\
\noalign{\smallskip}
HS 1946+7658     & IRAM  & 20 05 98$^{4}$ & 1300 & $<$10.4      \\
                 &       & 20 05 98$^{4}$ & 3000 & $<$5.0       \\
\noalign{\smallskip}
PG 2214+139      & IRAC1 & 21 06 96        &  4.7 & 44.2$\pm$4.6 \\
\noalign{\smallskip}
PG 2308+098      & IRAC1 & 21 06 96        &  3.7 & $<$17.56     \\
\noalign{\smallskip}
\hline    
\end{tabular}\\
$^{\dagger}$ Upper limits to the flux are given at the 3$\sigma$ level.
$^{1}$ Also observed on July 17, and 22.  
$^{2}$ Only four antenna were used on July 17. 
$^{3}$ Only three antenna were used on July 14, and four on July 17,
and 22.
$^{4}$ Also observed on May 23.
   \end{table}
%

   \begin{table}
      \caption[]{List of references of selected published data}
         \label{list_of_ref}
\begin{tabular}{@{}l c l@{}}
\hline
\noalign{\smallskip}
\multicolumn{1}{c}{Source Name} &
\multicolumn{1}{c}{\hspace*{1.0cm}}&
\multicolumn{1}{c}{References number$^{\dagger}$}\\ 
\noalign{\smallskip}
\hline
\noalign{\smallskip}
3C 47         &   &   1, 2, 3, 4, 5 \\
PKS 0135$-$247&   &   1, 3, 5, 6, 7 \\
PKS 0408$-$65 &   &   1 \\
PKS 0637$-$75 &   &   1, 2, 3, 5, 6, 8, 9 \\
PG 1004+130   &   &   1, 10, 11, 12, 13, 14, 15, 16 \\
PG 1048$-$090 &   &   1, 3, 11, 14, 15, 17 \\
4C 61.20      &   &   1, 3, 18 \\
PG 1100+772   &   &   1, 2, 3, 12, 15, 17, 19, 20, 21 \\ 
PG 1103$-$006 &   &   1, 12, 14, 15, 22, 23 \\
PG 1216+069   &   &   2, 12, 14, 23, 24, 25, 26 \\
PG 1352+183   &   &   2, 14, 15, 17, 25 \\
PG 1354+213   &   &   2, 14, 15, 25 \\
PG 1435$-$067 &   &   2, 14, 15, 25 \\
PG 1519+226   &   &   2, 14, 15, 25 \\
PG 1543+489   &   &   12, 14, 15, 25, 27, 28, 29 \\
PG 1718+481   &   &   1, 2, 14, 23, 26, 30 \\
B2 1721+34    &   &   1, 13, 17, 31 \\
HS 1946+7658  &   &   25, 32 \\
3C 405        &   &   1, 33, 34 \\
B2 2201+31A   &   &   1, 2, 7, 13, 16, 20, 31, 35, 36, 37\\
              &   &   38, 39, 40, 41 \\
PG 2214+139   &   &   1, 12, 15, 29, 30, 42 \\
PG 2308+098   &   &   1, 15, 30 \\
\noalign{\smallskip}
\hline    
\end{tabular}\\
\begin{tabular}{@{}l@{}l@{}}
$^{\dagger}$ 1:~NED; 2:~\cite{Gezari97};$\,\;$ & 3:~\cite{Brinkmann97};\\
4:~\cite{vanBemmel98};  & 5:~\cite{Kuhr81};   \\ 
6:~\cite{Tornikoski96}; & 7:~\cite{Steppe93}; \\
8:~\cite{Veron}; & 9:~\cite{Tanner96};\\
10:~\cite{Wilkes94}; & 11:~\cite{Kapahi95}; \\
12:~\cite{Sanders89}; & 13:~\cite{Lister94}; \\
14:~\cite{Neugebauer87}; & 15:~\cite{Miller93};\\
16:~\cite{Ennis82}; & 17:~\cite{Elvis94}; \\
18:~\cite{Reid95}; & 19:~\cite{ABA90};\\
20:~\cite{Chini89a}; & 21:~\cite{Lonsdale83}; \\
22:~\cite{Siebert98}; & 23:~\cite{Kellermann89}; \\
24:~\cite{Blundell98}; & 25:~\cite{Yuan98};\\
26:~\cite{Falcke96}; & 27:~\cite{Andreani99}; \\
28:~\cite{BLA96}; & 29:~\cite{CKB89b}; \\
30:~\cite{Wang96}; & 31:~\cite{Schartel96};\\
32:~\cite{Kuhn95}; & 33:~\cite{Robson98}; \\
34:~\cite{Haas98}; & 35:~\cite{Bloom99}; \\
36:~\cite{Hoekstra97}; & 37:~\cite{Mitchell94};\\
38:~\cite{Neugebauer86}; & 39:~\cite{Ghosh94}; \\
40:~\cite{Neugebauer79}; & 41:~\cite{Terasranta92}; \\
42:~\cite{Hughes93}. & \\
\end{tabular}
\end{table}

\section{ISOPHOT observations and data reduction}
\label{datared}

Photometric data at several (up to 11) wavelengths between 3.6 and 200
$\mu$m were obtained for each object using the single-element P1 and P2
detectors plus the two array cameras, C100 (3 pixels $\times$ 3 pixels) and
C200 (2 pixels $\times$ 2 pixels). Detector and observing parameters are
listed in Table~\ref{DetProp}. Most of the observations (124 in total for 16
objects) were performed in chopper mode, and the remaining (37 for 10
objects) by mapping the region surrounding the target (scans or rasters).

   \begin{table}
      \caption[]{ISOPHOT detector properties and covered sky region in
mapping observations}
         \label{DetProp}
\begin{tabular}{@{}l l c c c@{}}
\hline
\noalign{\smallskip}
\multicolumn{1}{l}{Detector} &
\multicolumn{1}{c}{$\lambda$}&
\multicolumn{1}{c}{Pixel}&
\multicolumn{1}{c}{Scan}&
\multicolumn{1}{c}{Raster}\\ 
\multicolumn{1}{l}{Name} &
\multicolumn{1}{c}{( $\mu$m )}&
\multicolumn{1}{c}{size}&
\multicolumn{2}{c}{Coverage}\\
\multicolumn{1}{l}{} &
\multicolumn{1}{c}{}&
\multicolumn{1}{c}{( $\arcsec$ )}&
\multicolumn{1}{c}{( $\arcsec\times\arcsec$ )}&
\multicolumn{1}{c}{( $\arcsec\times\arcsec$ )}\\ 
\noalign{\smallskip}
\hline
\noalign{\smallskip}
P1    & 3.6, 4.8, 7.3, 12 &  -   &  52$\times$156 &       -         \\
P2    & 25                &  -   &       -        &       -         \\
C100  & 60, 80, 100       & 43.5 & 138$\times$230 & 230$\times$230  \\
C200  & 150, 170, 200     & 89.4 & 184$\times$460 & 276$\times$460  \\
\noalign{\smallskip}
\hline    
\end{tabular}\\
\end{table}

In chopper mode the radiation beam is deflected from the source (on-source
position) to adjacent fields on the sky (off-source position) several times
in order to measure the background emission. Triangular (T) and rectangular
(R) chopping modes were used. In the triangular chopper mode the background
emission is measured in two different regions, while in the rectangular
chopper mode it is measured in only one position. Observing dates, filters,
apertures, exposure times, chopping mode, and measured fluxes are reported
in Table~\ref{DetObsChop} for each ISOPHOT chopper observations.

In mapping mode the telescope moves in a pattern around the source,
providing more sky coverage than in the chopper mode (Table~\ref{DetProp}).
P1 detector maps were performed with an aperture of 52$\arcsec$ during all
observations, except one (B2 2201+31A) during which the chosen aperture was
23$\arcsec$. Observing dates, filters, exposure times, and measured fluxes
are reported in Table~\ref{DetObsRast}. More details on mapping mode are
reported in section~\ref{BackSub}.

\subsection{First steps of the data reduction: from ERD to AAP level}
\label{erd_app}
The first part of the data reduction was performed using version 8.1 of the
PHT Interactive Analysis (PIA)\footnote{PIA is a joint development by the
ESA Astrophysics Division and the ISOPHOT consortium.} tool
(\cite{Gabriel97}). We started the reduction with the raw data processed
with version 8.7 of the Off-Line Processing (\cite{Laureijs98}). The raw
data form a sequence of detector read-outs distributed in 2$^{n}$ ($n$=2-6)
sets of four response curves or ramps, as function of time (Edited Raw Data:
ERD in Volts).

Each set of four ramps represents a sky position. Each ramp is corrected for
the non-linearity of the detector response, and for contamination of cosmic
particle events (glitches). The removal of read-outs affected by glitches is
carried out by applying two median filtering techniques: the
single-threshold technique that uses a threshold of 4.5 standard deviations
($\sigma$) for flagging bad read-outs and the two-threshold technique that
uses a threshold of 3.0$\sigma$ for flagging and 1.0$\sigma$ for
re-accepting read-outs. After applying the non-linearity correction and the
deglitching to the ERD, a straight line is fitted to each ramp, in order to
determine its slope or Signal per Ramp Data (SRD in Volt/s).

In most of the cases the first 25 or 50\% (1 or 2 ramps of 4) of the signals
per chopper plateau at the SRD level are discarded to enable the detector
response to stabilize at the level corresponding to the source flux density.
The remaining data are further corrected for highly discrepant points (value
at more than 3$\sigma$ from the average signal) still contaminated by
glitches, for the orbital dependent dark current, and for the signal
dependence on the ramp integration time (reset time interval) to obtain an
average Signal per Chopper Plateau (SCP in Volt/s).

After applying flat-fielding correction using PIA values, the SCP data are
calibrated to obtain the Standard Processed Data (SPD in unit of Watts).
Since the detector response varies with time, it is determined at the time
of the observation by measuring the flux emitted by two thermal Fine
Calibration Sources (FCS1 and FCS2) on board. The FCS measurements are
reduced in the same way as the scientific measurements up to this step. Data
from FCS1 are used because they are the best calibrated. The FCS1 signal is
checked in order to remove data with large uncertainties (this step is
equivalent to computing the weighted mean of the FCS1 data).

In the case of mapping observations, the FCS1 is observed twice, before and
after the observation of the source. The photometric calibration we use is
the average value of the two FCS1 measurements.

After the flux calibration the AAP (Auto Analysis Product) data are
obtained. They are a sequence of 2$^{n}$ off- and on-source flux
measurements (in Jy) each corresponding to a sky position. The reduction
from the AAP level to the final results is performed using our own IDL
routines, and not following the standard pipeline. This procedure was
also applied in the reduction of ISOPHOT chopper data of a sample of Seyfert
galaxies (\cite{Polletta99}).

   \begin{table}
      \caption[]{Details of ISOPHOT Observations performed in Chopper Mode}
         \label{DetObsChop}   
\begin{tabular}{@{}r c r c r r r@{}}
\hline
\noalign{\smallskip}
\multicolumn{1}{c}{$\lambda$} &
\multicolumn{1}{c}{Apt.} &
\multicolumn{1}{c}{Exp.} &
\multicolumn{1}{c}{Chop.} &
\multicolumn{1}{c}{F$_{\nu}$}&
\multicolumn{2}{c}{Uncert.$^{\dagger}$}\\
\multicolumn{1}{c}{($\mu$m)} &
\multicolumn{1}{c}{($\arcsec$)} &
\multicolumn{1}{c}{(s)} &
\multicolumn{1}{c}{Mode}&
\multicolumn{1}{c}{(mJy)} &
\multicolumn{1}{c}{Stat.}&
\multicolumn{1}{c}{Syst.}\\
\noalign{\smallskip}
\hline
\noalign{\smallskip}
\multicolumn{7}{c}{3C 47 (January 30, 1997)}\\
\hline
\noalign{\smallskip}
  4.8    &   23  &  128 & R &  $<$64. &        &     \\
 12.8    &   23  &  128 & R &     69. &  10.   & 21. \\
   20    &   23  &  128 & R & $<$171. &        &     \\
   60    &    -  &  256 & R &    170. &  52.   & 51. \\
  100    &    -  &  256 & R &    246. &  52.   & 74. \\
\noalign{\smallskip}
\hline
\noalign{\smallskip}
\multicolumn{7}{c}{PKS 0408$-$65 (June, 05 1997)}\\
\hline
\noalign{\smallskip}
  4.8    &   23  &  128 & R &  $<$26. &         &    \\
 12.8    &   23  &  128 & R &  $<$60. &         &    \\
   20    &   23  &  128 & R &  $<$84. &         &    \\
   60    &    -  &  128 & R & $<$308. &         &    \\
  100    &    -  &  128 & R & $<$162. &         &    \\
\noalign{\smallskip}
\hline
\noalign{\smallskip}
\multicolumn{7}{c}{PKS 0637$-$75 (June, 05 1997)}\\
\hline
\noalign{\smallskip}
  4.8    &   23  &  128 & R &  $<$26. &         &    \\
 12.8    &   23  &  128 & R &  $<$43. &         &    \\
   20    &   23  &  128 & R &  $<$99. &         &    \\
   60    &    -  &  128 & R & $<$117. &         &    \\
  100    &    -  &  128 & R & $<$143. &         &    \\
\noalign{\smallskip}
\hline
\noalign{\smallskip}
\multicolumn{7}{c}{4C 61.20 (April, 27 1996)}\\
\hline
\noalign{\smallskip}
  7.3    & 13.8  &  512 & T &  $<$35. &         &    \\
   12    &   23  &  256 & T &  $<$34. &         &    \\
   25    &   52  &  512 & T & $<$110. &         &    \\
   60    &    -  &  128 & R & $<$278. &         &    \\
   80    &    -  &  128 & R &    258. &  48.    & 77.\\
  100    &    -  &  128 & R & $<$303. &         &    \\
  150    &    -  &  128 & R & $<$267. &         &    \\
  170    &    -  &  128 & R & $<$148. &         &    \\
  200    &    -  &  128 & R & $<$335. &         &    \\
\noalign{\smallskip}  
\hline
\end{tabular}
\begin{list}{}{}
\item[$^{\dagger}$] In units of mJy.
\end{list}
\end{table}

\begin{table}
      \setcounter{table}{4}%
      \caption[]{(continued)}
\begin{tabular}{@{}r c r c r r r@{}}
\hline
\noalign{\smallskip}
\multicolumn{1}{c}{$\lambda$} &
\multicolumn{1}{c}{Apt.} &
\multicolumn{1}{c}{Exp.} &
\multicolumn{1}{c}{Chop.} &
\multicolumn{1}{c}{F$_{\nu}$}&
\multicolumn{2}{c}{Uncert.$^{\dagger}$}\\
\multicolumn{1}{c}{($\mu$m)} &
\multicolumn{1}{c}{($\arcsec$)} &
\multicolumn{1}{c}{(s)} &
\multicolumn{1}{c}{Mode}&
\multicolumn{1}{c}{(mJy)} &
\multicolumn{1}{c}{Stat.}&
\multicolumn{1}{c}{Syst.}\\
\noalign{\smallskip}
\hline
\noalign{\smallskip}
\multicolumn{7}{c}{PG 1100+772 (June, 17 1996)}\\
\hline
\noalign{\smallskip}
  4.8    &   52  &  512 & R &     25. &   7.    &  8. \\
  7.3    &   52  &  512 & R &     32. &   5.    & 10. \\
   12    &   52  &  512 & R &     36. &   8.    & 11.  \\
   25    &  120  &  512 & R &     54. &  10.    & 16. \\
   60    &    -  &  256 & R &     70. &  14.    & 21. \\
  100    &    -  &  128 & R & $<$282. &         &     \\
  150    &    -  &  512 & R & $<$417. &         &     \\
  200    &    -  &  512 & R & $<$468. &         &     \\
\noalign{\smallskip}  
\hline
\noalign{\smallskip}  
\multicolumn{7}{c}{PG 1216+069 (July, 11 1996)}\\
\hline
\noalign{\smallskip}
  4.8    &   52  &  512 & R &     50. &   15.   & 15.\\   
  7.3    &   52  &  512 & R &     72. &    6.   & 22.\\   
   12    &   52  &  512 & R &     90. &   12.   & 27.\\   
   25    &  120  &  512 & R &     95. &   26.   & 29.\\   
   60    &   -   &  128 & R &     65. &   14.   & 20.\\   
  100    &   -   &  128 & R & $<$140. &         &    \\   
  150    &   -   &  512 & R & $<$137. &         &    \\   
  200    &   -   &  512 & R & $<$149. &         &    \\   
\noalign{\smallskip}  
\hline
\noalign{\smallskip}  
\multicolumn{7}{c}{PG 1352+183 (December, 14 1996)}\\
\hline
\noalign{\smallskip}
   60    &   -   &   64 & R &    198. &   23.   & 59.\\   
   80    &   -   &   64 & R & $<$237. &         &    \\   
  100    &   -   &   64 & R & $<$213. &         &    \\   
  150    &   -   &   64 & R & $<$283. &         &    \\   
  170    &   -   &   64 & R & $<$240. &         &    \\   
  200    &   -   &   64 & R & $<$206. &         &    \\   
\noalign{\smallskip}  
\hline
\noalign{\smallskip}  
\multicolumn{7}{c}{PG 1354+213 (June, 13 1996)}\\
\hline
\noalign{\smallskip}
  7.3    & 13.8  & 2048 & T &     18. &   5.   &  5.\\   
   12    &   23  & 1024 & T &     21. &   5.   &  6.\\   
   25    &   52  & 2048 & T &     23. &   7.   &  7.\\   
   60    &    -  &  128 & R & $<$298. &        &    \\   
   80    &    -  &  128 & R & $<$194. &        &    \\   
  100    &    -  &  128 & R & $<$125. &        &    \\   
  150    &    -  &  128 & R & $<$170. &        &    \\   
  170    &    -  &  128 & R & $<$137. &        &    \\   
  200    &    -  &  128 & R & $<$280. &        &    \\   
\noalign{\smallskip}  
\hline
\end{tabular}
\begin{list}{}{}
\item[$^{\dagger}$] In units of mJy.
\end{list}
\end{table}

\begin{table}
      \setcounter{table}{4}%
      \caption[]{(continued)}
\begin{tabular}{@{}r c r c r r r@{}}
\hline
\noalign{\smallskip}
\multicolumn{1}{c}{$\lambda$} &
\multicolumn{1}{c}{Apt.} &
\multicolumn{1}{c}{Exp.} &
\multicolumn{1}{c}{Chop.} &
\multicolumn{1}{c}{F$_{\nu}$}&
\multicolumn{2}{c}{Uncert.$^{\dagger}$}\\
\multicolumn{1}{c}{($\mu$m)} &
\multicolumn{1}{c}{($\arcsec$)} &
\multicolumn{1}{c}{(s)} &
\multicolumn{1}{c}{Mode}&
\multicolumn{1}{c}{(mJy)} &
\multicolumn{1}{c}{Stat.}&
\multicolumn{1}{c}{Syst.}\\
\noalign{\smallskip}  
\hline
\noalign{\smallskip}  
\multicolumn{7}{c}{PG 1435$-$067 (January, 07 1997)}\\
\hline
\noalign{\smallskip}
  7.3    & 13.8  &  256 & T &     70. &   9.   & 21.\\
   12    &   23  &   64 & T &  $<$94. &        &    \\
   25    &   52  &  128 & T & $<$718. &        &    \\
   60    &    -  &   64 & R &$<$1111. &        &    \\
   80    &    -  &   64 & R & $<$849. &        &    \\
  100    &    -  &   64 & R & $<$333. &        &    \\
  150    &    -  &   64 & R & $<$386. &        &    \\
  170    &    -  &   64 & R & $<$369. &        &    \\
  200    &    -  &   64 & R & $<$317. &        &    \\
\noalign{\smallskip}  
\hline
\noalign{\smallskip}  
\multicolumn{7}{c}{PG 1519+226 (February, 01 1997)}\\
\hline
\noalign{\smallskip}
  7.3    & 13.8  &  256 & T &     32. &    8.  & 10.\\
   12    &   23  &   64 & T &  $<$48. &        &    \\
   25    &   52  &   64 & T &     76. &   17.  & 23.\\
   60    &    -  &  128 & R &    172. &   44.  & 52.\\
   80    &    -  &  128 & R & $<$194. &        &    \\
  100    &    -  &  128 & R &    121. &   22.  & 36.\\
  150    &    -  &  128 & R & $<$207. &        &    \\
  170    &    -  &  128 & R & $<$151. &        &    \\
  200    &    -  &  128 & R & $<$154. &        &    \\
\noalign{\smallskip}
\hline
\noalign{\smallskip}
\multicolumn{7}{c}{PG 1543+489 (May, 30 1996)}\\
\hline
\noalign{\smallskip}
  4.8    &   52  &  512 & R &     20. &    6.  &  6.\\
  7.3    &   52  &  256 & R &     39. &    4.  & 12.\\
   12    &   52  &  256 & R &     45. &    6.  & 14.\\
   25    &  120  &  512 & R &    140. &   13.  & 42.\\
   60    &    -  &  128 & R &    470. &  144.  &141.\\
  100    &    -  &  128 & R &    399. &   55.  &120.\\
  150    &    -  &  128 & R &    455. &  117.  &182.\\
  200    &    -  &  256 & R & $<$377. &        &    \\
\noalign{\smallskip}
\hline
\noalign{\smallskip}  
\multicolumn{7}{c}{PG 1718+481 (May, 30 1996)}\\
\hline
\noalign{\smallskip}
  4.8    &   52  &  512 & R &     18. &  3.  &  5.\\
  7.3    &   52  &  256 & R &     24. &  6.  &  7.\\
   12    &   52  &  256 & R &     25. &  6.  &  8.\\
   25    &  120  &  512 & R &     44. &  7.  & 13.\\
   60    &    -  &  128 & R &     90. &  1.  & 27.\\
  100    &    -  &  128 & R &     56. &  1.  & 17.\\
  150    &    -  &  128 & R & $<$251. &      &    \\
  200    &    -  &  256 & R & $<$261. &      &    \\
\noalign{\smallskip}  
\hline
\end{tabular}
\begin{list}{}{}
\item[$^{\dagger}$] In units of mJy.
\end{list}
\end{table}

   \begin{table}
      \setcounter{table}{4}%
      \caption[]{(continued)}
\begin{tabular}{@{}r c r c r r r@{}}
\hline
\noalign{\smallskip}
\multicolumn{1}{c}{$\lambda$} &
\multicolumn{1}{c}{Apt.} &
\multicolumn{1}{c}{Exp.} &
\multicolumn{1}{c}{Chop.} &
\multicolumn{1}{c}{F$_{\nu}$}&
\multicolumn{2}{c}{Uncert.$^{\dagger}$}\\
\multicolumn{1}{c}{($\mu$m)} &
\multicolumn{1}{c}{($\arcsec$)} &
\multicolumn{1}{c}{(s)} &
\multicolumn{1}{c}{Mode}&
\multicolumn{1}{c}{(mJy)} &
\multicolumn{1}{c}{Stat.}&
\multicolumn{1}{c}{Syst.}\\
\noalign{\smallskip}  
\hline
\noalign{\smallskip}
\multicolumn{7}{c}{B2 1721+34 (April, 20 1996)}\\
\hline
\noalign{\smallskip}
  7.3    & 13.8  &  512 & T &    19.$^{\ddagger}$  &  5.  &  6.\\
   12    &   23  &  512 & T & $<$26.$^{\ddagger}$  &      &    \\
   25    &   52  &  512 & T &$<$135.$^{\ddagger}$  &      &    \\
\noalign{\smallskip}  
\hline
\noalign{\smallskip}  
\multicolumn{7}{c}{HS 1946+7658 (May, 27 1996)}\\
\hline
\noalign{\smallskip}
  4.8    &   52  & 1024 & R &   $<$5. &      &    \\
  7.3    &   52  & 1024 & R &      4.5&  1.2 & 1.4\\
   12    &   52  & 1024 & R &      8.7&  0.7 & 2.6\\
   25    &  120  & 1024 & R &     47. & 15.  & 14.\\
   60    &    -  &   64 & R & $<$230. &      &    \\
  100    &    -  &   64 & R & $<$300. &      &    \\
  150    &    -  &   64 & R &    682.$^{\ddagger}$ & 17.  &273.\\
  200    &    -  &   64 & R &    378.$^{\ddagger}$ & 27.  &151.\\
\noalign{\smallskip}
\hline
\noalign{\smallskip}  
\multicolumn{7}{c}{B2 2201+31A (November, 16 1996)}\\
\hline
\noalign{\smallskip}
  3.6    &    5  & 2048 & T &  $<$17. &      &    \\
  4.8    &  7.6  & 1024 & T &  $<$52. &      &    \\
  7.3    & 13.8  &  256 & T &     45. &  5.  & 14.\\
   12    &   23  &   64 & T &     64. &  8.  & 19.\\
   12    &   23  &   64 & T &  $<$62. &      &    \\
   25    &   52  &  128 & T &    106. & 25.  & 32.\\
   60    &    -  &  128 & R &    193. & 13.  & 58.\\
   80    &    -  &  128 & R &    139. & 11.  & 42.\\
  100    &    -  &  128 & R &    212. & 51.  & 64.\\
  150    &    -  &  128 & R &    847. & 97.  &339.\\
  170    &    -  &  128 & R &    699. & 98.  &280.\\
  200    &    -  &  128 & R &    382. & 97.  &153.\\
\noalign{\smallskip}  
\hline
\noalign{\smallskip}  
\multicolumn{7}{c}{PG 2308+098 (November, 26 1996)}\\
\hline
\noalign{\smallskip}
  3.6    &    5  & 2048 & T &  $<$31. &     &   \\
  4.8    &  7.6  & 1024 & T &     13. & 2.  & 4.\\
   12    &   23  &  128 & T &  $<$39. &     &   \\
\noalign{\smallskip}
\hline
\noalign{\smallskip}
\multicolumn{7}{c}{PG 2308+098 (December, 12 1996)}\\
\hline
\noalign{\smallskip}
  7.3    & 13.8  &  256 & T &  $<$21. &     &   \\
   12    &   23  &  128 & T &     16. & 55. & 5.\\
   25    &   52  &  512 & T & $<$108. &     &   \\
   60    &    -  &  128 & R & $<$183. &     &   \\
   80    &    -  &  128 & R & $<$174. &     &   \\
  100    &    -  &  128 & R &  $<$83. &     &   \\
  150    &    -  &  128 & R & $<$201. &     &   \\
  170    &    -  &  128 & R & $<$152. &     &   \\
  200    &    -  &  128 & R & $<$214. &     &   \\
\noalign{\smallskip}  
\hline
\end{tabular}\\
$^{\dagger}$ In units of mJy.
$^{\ddagger}$ Doubtful data. Discussed in Sec.~\ref{syst_unc} and~\ref{nature}.
\end{table}

   \begin{table}
      \caption[]{Details of ISOPHOT Observations performed in Raster Mode}
         \label{DetObsRast}
\begin{tabular}{ @{}l r r r r r@{} }
\hline
\noalign{\smallskip}
\multicolumn{1}{c}{Obs. date} &
\multicolumn{1}{c}{$\lambda$} &
\multicolumn{1}{c}{Exp.} &
\multicolumn{1}{c}{F$_{\nu}$}&
\multicolumn{2}{c}{Uncert.$^{\dagger}$}\\
\multicolumn{1}{c}{} &
\multicolumn{1}{c}{($\mu$m)} &
\multicolumn{1}{c}{(s)} &
\multicolumn{1}{c}{(mJy)}&
\multicolumn{1}{c}{Stat.}&
\multicolumn{1}{c}{Syst.}\\
\noalign{\smallskip}  
\hline
\noalign{\smallskip}
\multicolumn{6}{c}{3C 47}\\
\hline
\noalign{\smallskip}
January, 31 1998 &  60       & 636  & $<$129. &   &   \\
\noalign{\smallskip}
\hline
\noalign{\smallskip}
\multicolumn{6}{c}{PKS 0135$-$247}\\
\hline
\noalign{\smallskip}
December, 12 1997 &  12$^{d}$ &1206  & $<$199. &   &   \\
\noalign{\smallskip}
\hline
\noalign{\smallskip}
\multicolumn{6}{c}{PKS 0408$-$65}\\
\hline
\noalign{\smallskip}
June, 05 1997 & 170$^{a}$ & 195  &  $<$51. &   &   \\
\noalign{\smallskip}
\hline
\noalign{\smallskip}
\multicolumn{6}{c}{PKS 0637$-$75}\\
\hline
\noalign{\smallskip}
June, 05 1997 & 170$^{a}$ & 346  & $<$180. &     &     \\
March, 15 1998 & 150$^{a}$ & 387  & $<$171. &     &     \\
             & 200$^{a}$ & 387  & $<$203. &     &     \\
\noalign{\smallskip}
\hline
\noalign{\smallskip}
\multicolumn{6}{c}{PG 1100+772}\\
\hline
\noalign{\smallskip}
August, 15 1997 &  60         & 348  &  $<$99. &     &     \\
             & 100         & 348  &  $<$55. &     &     \\
             & 150$^{b,c}$ & 579  &  $<$82. &     &     \\
             & 200$^{b,c}$ & 579  &  $<$54. &     &     \\
October, 28 1997 &  12$^{d}$   & 954  & $<$161. &     &     \\
             &  60$^{d}$   & 630  &     68. & 22. & 20. \\
November, 02 1997 & 150$^{a}$   & 579  & $<$120. &     &     \\
\noalign{\smallskip}
\hline
\noalign{\smallskip}
\multicolumn{6}{c}{PG 1543+489}\\
\hline
\noalign{\smallskip}
November, 01 1997 &  12$^{d}$   & 864  & $<$202. &      &    \\
             &  60$^{d}$   & 630  &    269. &  27. & 81.\\
             & 150$^{a,d}$ & 574  &    279. &  45. & 84.\\
\noalign{\smallskip}  
\hline
\end{tabular}\\
$^{\dagger}$ In units of mJy. $^{a}$4$\times$2 map. $^{b}$2$\times$4 map.
$^{c}$Step amplitude of 90$\arcsec$. $^{d}$Telescope nodding mode.
\end{table}

   \begin{table}
      \setcounter{table}{5}%
      \caption[]{(continued)}
\begin{tabular}{ @{}l r r r r r@{} }
\hline
\noalign{\smallskip}
\multicolumn{1}{c}{Obs. date} &
\multicolumn{1}{c}{$\lambda$} &
\multicolumn{1}{c}{Exp.} &
\multicolumn{1}{c}{F$_{\nu}$}&
\multicolumn{2}{c}{Uncert.$^{\dagger}$}\\
\multicolumn{1}{c}{} &
\multicolumn{1}{c}{($\mu$m)} &
\multicolumn{1}{c}{(s)} &
\multicolumn{1}{c}{(mJy)}&
\multicolumn{1}{c}{Stat.}&
\multicolumn{1}{c}{Syst.}\\
\noalign{\smallskip}  
\hline
\noalign{\smallskip}
\multicolumn{6}{c}{PG 1718+481}\\
\hline
\noalign{\smallskip}
April, 20 1997 &  60         & 348  &     59. &  16. & 18.\\
             & 100         & 348  &     40. &  13. & 12.\\
             & 150$^{a}$   & 323  & $<$104. &      &    \\
             & 200$^{a}$   & 579  &  $<$56. &      &    \\
November, 01 1997 &  12$^{d}$   & 864  & $<$164. &      &    \\
             &  60$^{d}$   & 630  &     69. &  13. & 21.\\
             & 150$^{a,d}$ & 574  & $<$145. &      &    \\
\noalign{\smallskip}
\hline
\noalign{\smallskip}
\multicolumn{6}{c}{HS 1946+7658}\\
\hline
\noalign{\smallskip}
April, 28 1997 &  60       & 348  &     49. &  12. & 15.\\
             & 100       & 348  &     79. &  14. & 24.\\
             & 150$^{a}$ & 451  & $<$114. &      &    \\
             & 200$^{a}$ & 579  & $<$356. &      &    \\
October, 26 1997 & 200$^{a}$ & 579  & $<$227. &      &    \\
November, 17 1997 &  12$^{d}$ & 954  & $<$171. &      &    \\
November, 18 1997 &  12$^{d}$ & 954  & $<$168. &      &    \\
             &  60$^{d}$ & 630  &     66. &  15. & 20.\\
December, 08 1997 &  60$^{d}$ & 630  &     68. &  20. & 20.\\
\noalign{\smallskip}
\hline
\noalign{\smallskip}
\multicolumn{6}{c}{3C 405}\\
\hline
\noalign{\smallskip}
October, 30 1997 &  60$^{d}$   & 630  &  2360. &  76. & 708.\\
             & 170$^{a,d}$ & 574  &  1582. & 160. & 475.\\
\noalign{\smallskip}
\hline
\noalign{\smallskip}
\multicolumn{6}{c}{B2 2201+31A}\\
\hline
\noalign{\smallskip}
November, 20 1997  &  12$^{d}$ & 852  &    55. &  15. & 17.\\
	      &  60$^{d}$ & 414  &   111. &  20. & 33.\\
	      & 170$^{a}$ & 387  &$<$249. &      &    \\
\noalign{\smallskip}  
\hline
\end{tabular}\\
$^{\dagger}$ In units of mJy. $^{a}$4$\times$2 map. $^{b}$2$\times$4 map.
$^{c}$Step amplitude of 90$\arcsec$. $^{d}$Telescope nodding mode.
\end{table}

\subsection{From the AAP level to final results}
\label{aap_final}
The last steps of the data reduction before determining the source flux are
the background subtraction, the deletion of remaining highly discrepant
points, and the correction for effects depending on chopper plateau time,
vignetting (only for chopper observations) and point spread function.

In the case of chopper observations with the C100 detector only the central
pixel pointed on the source is considered to derive the flux density, since
the eight border pixels contain only a small fraction of the central point
source and summing these values would hence only increase the noise.

\subsubsection{Background subtraction}
\label{BackSub}

In chopper observations the background is measured at each off-source
position. Since in some cases the instruments show long term drift effects,
the background signal is estimated near the time of each on-source
measurement and subtracted. In the case of chopper observations the
background estimates are obtained by computing the weighted mean of each
pair of consecutive off-source measurements. The weights are computed from
PIA statistical uncertainties. Since the sequence of chopper plateaux ends
with an on-source position, we used the weighted mean of the two last
on-source measurements, and the flux observed in the last off-source
position to determine the last pair of on- and off-source values, for a
total of 2$^{n-1}$ flux values.

\begin{figure}
     \resizebox{8.8cm}{!}{\includegraphics{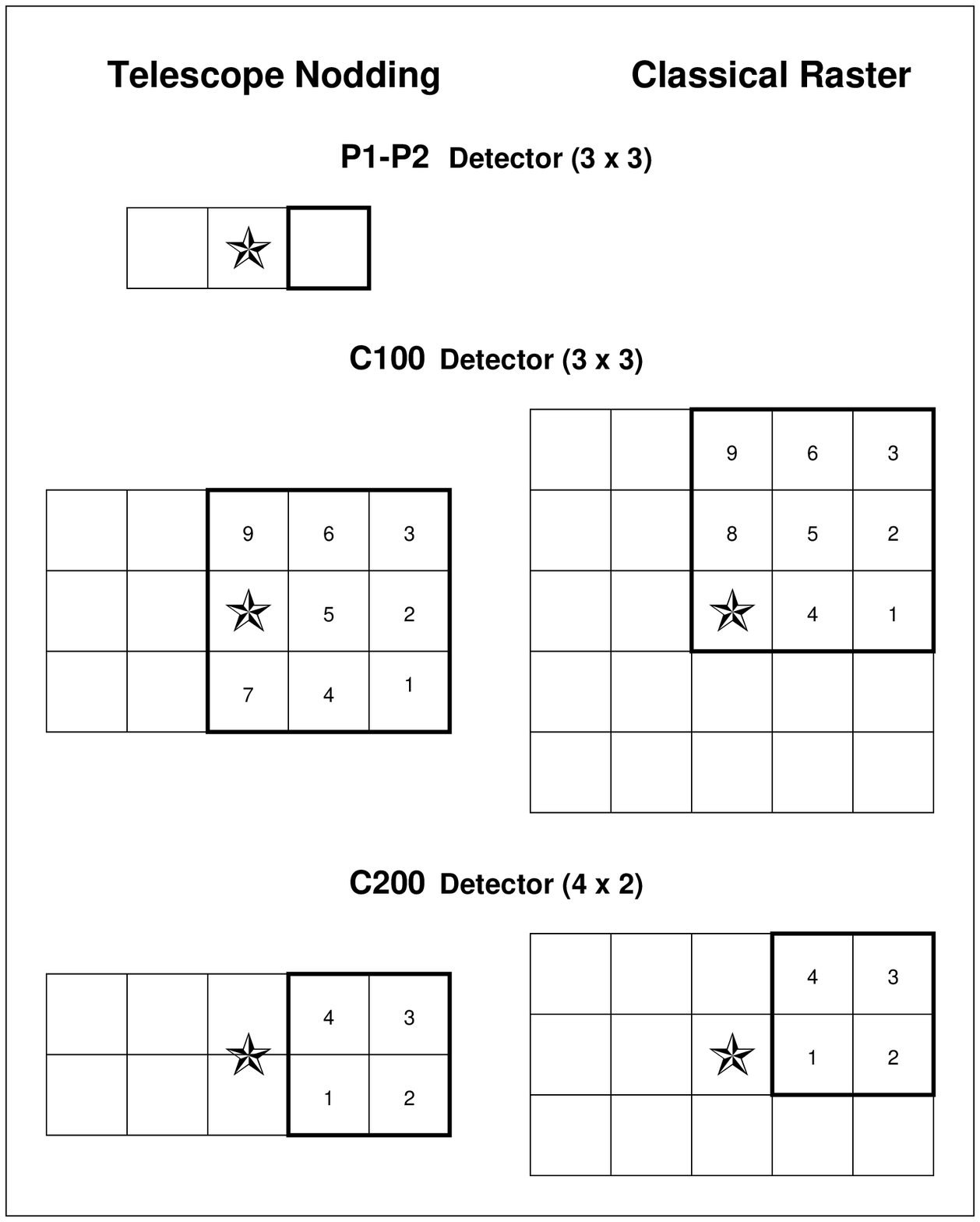}}
     \caption[]{Schematic representation of the two observing modes used for
      mapping observations: telescope nodding (scan) on the left side and
      classical raster on the right side. The position of the source in the
      different cases is represented by a star. The bold square represents
      the initial position of the detector. The numbering corresponds to the
      pixels numbers. The detectors move first from the left to the right,
      then in the opposite direction. In the case of classical raster, at
      each change of the horizontal direction, the detector shifts down one
      step.  All the steps are separated by the pixel size. The reported
      pixel sizes do not correspond to the real proportions. The numbers in
      parenthesis refer to the number of horizontal steps $\times$ the
      number of repetitions for nodding and vertical steps of telescope
      motion in the raster maps.}
        \label{Figrastmod}
 \end{figure}

Small maps of the regions immediately surrounding ten of the targets were
constructed in one or both of the following ways (Fig.~\ref{Figrastmod}):
multiple linear scans across the source and rastering the detector about the
source. A scan with the P1 and C100 detectors consisted of three steps of
the telescope, with this sequence repeated three times.  Only the middle row
of C100 pixels (8, 5, and 2, as depicted in Fig.~\ref{Figrastmod}) viewed
the source. The C200 scans contained four steps, repeated twice, with the
source centered between two pixels. Note that the source was observed in
only the middle two steps of the C200 scan. The raster patterns were
3$\times$3, 3$\times$3, and 4$\times$2 or 2$\times$4, for the P1, C100, and
C200 detectors, respectively.  Each pixel viewed the source once in the
raster maps.  The step size between exposures for both scans and rasters was
approximately equal to the pixel size, a little more to take into account
the gap between pixels, for the C100 and C200 cameras, and equal to the
aperture size for the P1 detectors, which resulted in a different total sky
coverage for each detector (see Table~\ref{DetProp}). A background estimate
for each on-source measurement was obtained from a weighted average of the
flux measured in the raster or scan positions immediately preceding and
following the source position by the same pixel. Using the same pixel to
determine the background reduced the impact of uncertainties in the flat
field. The weighted average background was subtracted from each on-source
measurement providing a sequence of source flux values.

Residual effects of detector instabilities, in both chopper observations and
maps, produced occasional discrepant points, which were culled by one-pass
sigma clipping. The threshold number of standard deviations to reject a flux
value depended on the number of points in the sequence (Chauvenet's
criterion in Taylor (1982)), ranging from 1.15$\sigma$ to 2.66$\sigma$. 

In the case of chopper observations with the C200 camera, the source flux
was computed by adding together the fluxes measured by each of the four
pixels. The source flux was divided between pairs of adjoining pixels in the
C200 scans; these were averaged by weighting with their uncertainties.
Table~\ref{DetObsRast} lists the weighted mean of each flux sequence after
clipping, or 3$\sigma$ upper limits for non--detection, where $\sigma$ is the
quadratic sum of the statistical and systematic uncertainties (see
section~\ref{syst_unc}), also reported in the tables.

\subsubsection{Vignetting correction}
\label{vign}

In the case of chopper observations with the C100 and C200 detectors, the
data are further corrected for the signal loss outside the beam of the
telescope (vignetting). The PIA default vignetting correction factors were
computed considering the dependency only on the distance of the chopper
positions and on the filter, but recent investigations have shown that they
depend also on the time per chopper plateau (M. Haas, private
communication). 

The PIA default vignetting factors were applied directly to the data,
resulting in little change to the flux values obtained with the C100 camera.
However, the fluxes from the C200 detector are $\sim$ 80\%, in median,
larger after the vignetting correction. Given the uncertainty over the
accuracy of the vignetting factors, we averaged the fluxes derived with and
without the correction and added the respective uncertainties in
quadrature.

\subsubsection{Correction for effects depending on the chopper plateau time and point spread function}
\label{trans_psf}
In the case of chopper observations a correction for the signal loss due to
effects depending on the time per chopper plateau is applied to
the computed flux value. Since short integration times with the C100 and
C200 detectors do not reach the full signal, the observed flux can be
reduced by large factors, typically up to 68 \% for the C100 detector, and
up to 12\% for the C200 detector for the shortest observations. The
correction factors we use are the PIA default values.

All the computed fluxes are finally corrected for the point spread function
(psf) of each detector using the default PIA values derived empirically in
most of the cases. The available psf correction values correspond to a
source position centered in a pixel or located in a corner of the pixel. In
case of scans with the C200 detector we needed the psf correction
corresponding to the target located in the middle of a side of the pixel. We
derived it by assuming a bi-dimensional Gaussian function for the psf and
constraining its parameters using the other two known values for each
wavelength (30.6\% at 150 $\mu$m, and 29.1\% at 170 $\mu$m).

\subsection{Calculation of systematic uncertainties}
\label{syst_unc}
During the data reduction only statistical errors were taken into account,
and not those in the absolute flux density calibration. The accuracy of the
absolute photometric calibration depends mainly on systematic errors
(detector transient effects, calibration response, dark current, point
spread function), and it is currently known to be better than 30\%
(\cite{Klaas98}). The associated statistical and systematic uncertainties
are reported in the last two columns of Table~\ref{DetObsChop}
and~\ref{DetObsRast}. We associate to the measured flux an uncertainty that
is the quadratic sum of their statistical uncertainties and 30\% (for the
C200 measurements obtained in chopper mode we use 40\% that corresponds to
half of the uncertainty due to the vignetting correction) of the measured
value. Among our ISOPHOT observations the datasets of B2 1721+34 were
identified as failed after inspection of their quality by the ISO team
because they were heavily affected by cosmic rays and thus scientifically
useless. We report these data in Table~\ref{DetObsChop}, and indicate that
they are doubtful in a footnote.

\section{The spectral energy distributions}

We report the first far-IR detection for 9 (4C 61.20, PG 1216+069, PG 1352+183,
PG 1354+213, PG 1435$-$067, PG 1519+226, PG 1718+481, PG 2308+098, and
HS 1946+7658) of the 18 sources observed with ISOPHOT. Among the remaining 9
sources, three (PKS 0135$-$247, PKS 0408$-$65, and PKS 0637$-$75) were not
detected, and the remainings 6 were already detected by IRAS.

All the data were converted to monochromatic luminosities in the rest
frame of the object (H$_\mathrm{0}$ = 75 km s$^{-1}$ Mpc$^{-1}$,
q$_\mathrm{0}$=0.5) using the
following equation:
\begin{equation}
      L_{\nu\, \mathrm{em}} = 4 \pi d_\mathrm{L}^{2} F_{\nu\, \mathrm{obs}} / (1 + z),
\end{equation}
where $F_{\nu\, \mathrm{obs}}$ is the monochromatic flux in the observer's frame, and
$d_\mathrm{L}$ is the luminosity distance to the object.
In the case of PKS 0408$-$65 we adopted a redshift value equal to 0.5,
arbitrarily chosen, since no redshift measurement is available.  This source
will not be considered in the following analysis. 

The spectral energy distributions (SEDs) as $\nu L_{\nu}$ versus $\nu$ in
the rest frame of all sources are shown in Fig.~\ref{FigsedFSRQ}. 
Upper limits are plotted only if there is no detection at that frequency at
any epoch, and in cases of multiple upper limits at the same frequency from
different epochs, only the most stringent is used. A broad band spectrum of
a local galaxy in its rest frame, representing the AGN host galaxy, is
superposed in each panel of Fig.~\ref{FigsedFSRQ}. We chose the spiral
galaxy M100, modeled by Silva et~al. (1998), to which we added data in the
radio (\cite{Becker91};~\cite{Gregory91};~White \& Becker 1992) and in the
soft X-ray energy domains (\cite{Immler98}) in the case of RQQ, and the
template of a giant elliptical galaxy (\cite{Silva98}) in the case of RLQ,
RIQ and RG. The reported host galaxy template was not modified by any
normalization. In many cases this is orders of magnitude below the observed
luminosities, and even if it is shifted towards higher luminosities to reach
the quasar SED, it will remain below at most frequencies.
   \begin{figure*}
      \resizebox{14cm}{!}{\includegraphics{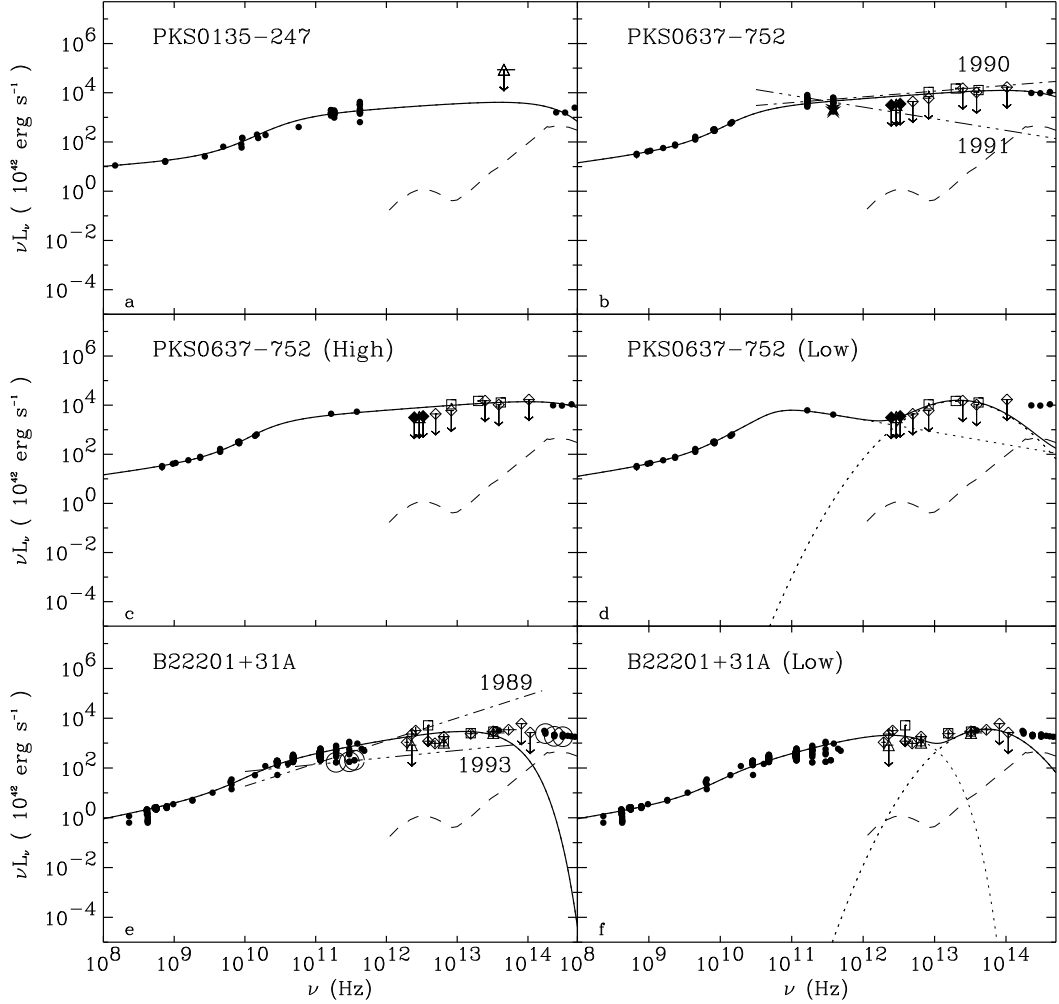}}
     \caption{SEDs as $\nu L_{\nu}$ versus $\nu$ in the rest frame of the
     objects. All sources shown on this page are FSRQ. Full circles
     represent literature data, stars SEST data, open diamonds ISOPHOT
     chopper data, triangles and filled diamonds ISOPHOT raster data at
     different epochs, and squares IRAS data. Arrows represent
     3$\sigma$ upper limits. The solid line represents the sum of all
     single fitted components represented by dotted lines. The dotted
     lines represent the best fit non-thermal models of the radio component
     (see section~\ref{rad_cont}, and Table~\ref{nonthermal_param}), and the
     parabolic fits of the IR component (see section~\ref{multibands}). When
     more than one radio component is present a symbol is reported for each
     spectral component: C for core and L for lobe (L1 and L2 for two
     lobes). A typical host galaxy template (dashed line), a spiral galaxy
     in the case of RQQ, and a giant elliptical galaxy in the case of RLQ,
     RIQ and RG, is over-plotted. The radio spectrum of PKS 0637$-$752 was
     fit in two ways, using only the simultaneous mm data corresponding to
     the flattest power law fit ({\bf c}) (dash-dot line in {\bf b}) and the
     two corresponding to the steepest power law fit ({\bf d})
     (dash-dot-dot-dot line in {\bf b}). The radio spectrum of B2 2201+31A
     was also fit in two ways, taking all data from radio to mid-IR ({\bf
     e}) and only data from radio to far-IR ({\bf f}). The flattest power
     law fit (dash-dot line in {\bf e}) and the steepest one
     (dash-dot-dot-dot in {\bf e}) measured during simultaneous mm
     observations of B2 2201+31A are reported. The big open circles in {\bf
     e} correspond to simultaneous observations at mm and near-IR
     wavelengths.}
              \label{FigsedFSRQ}
    \end{figure*}
%
   \begin{figure*}
      \setcounter{figure}{1}%
      \resizebox{14cm}{!}{\includegraphics{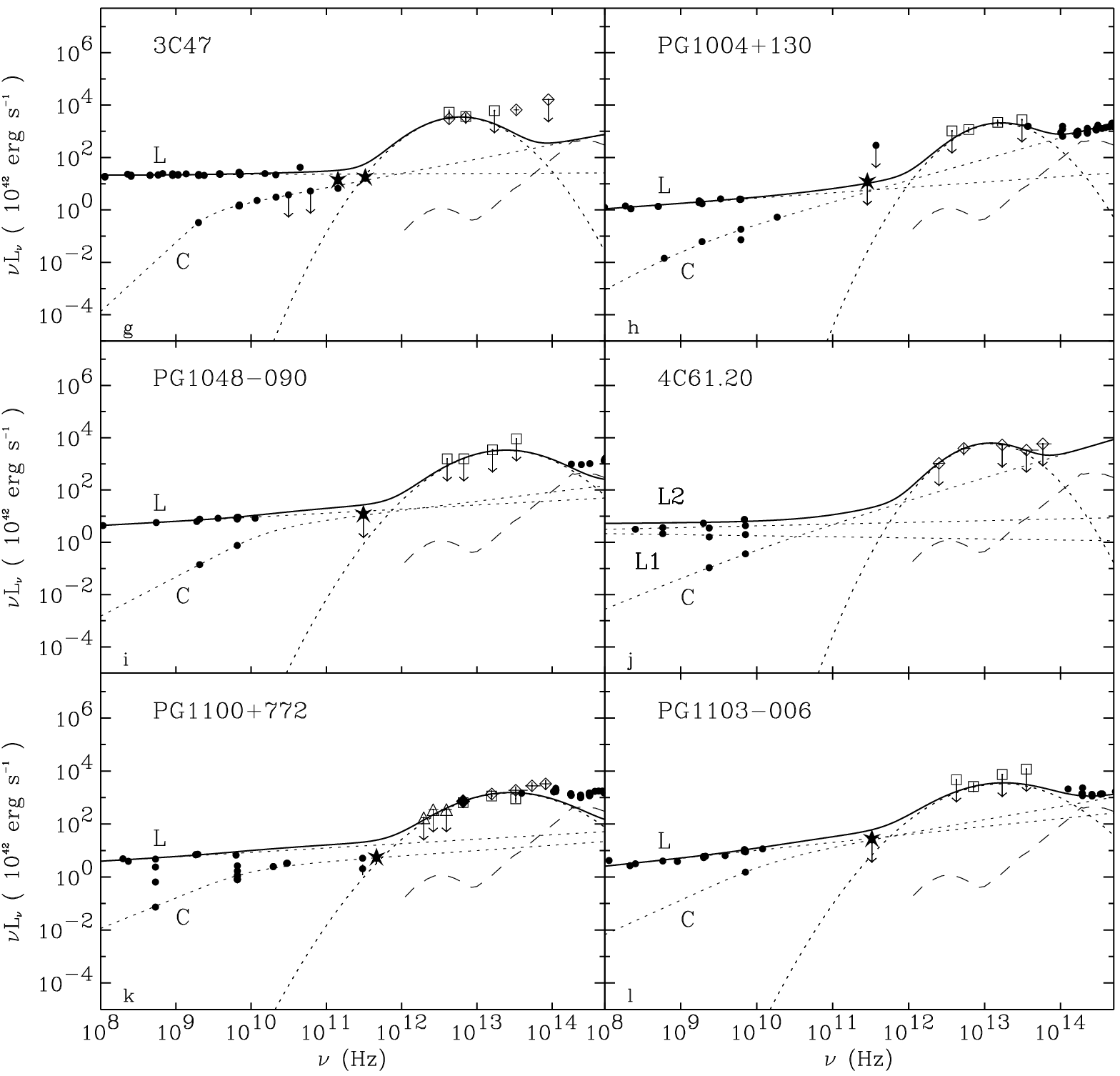}}
     \caption{Continued. SEDs for SSRQ are displayed on this page.}
    \end{figure*}
%
   \begin{figure*}
      \setcounter{figure}{1}%
      \resizebox{14cm}{!}{\includegraphics{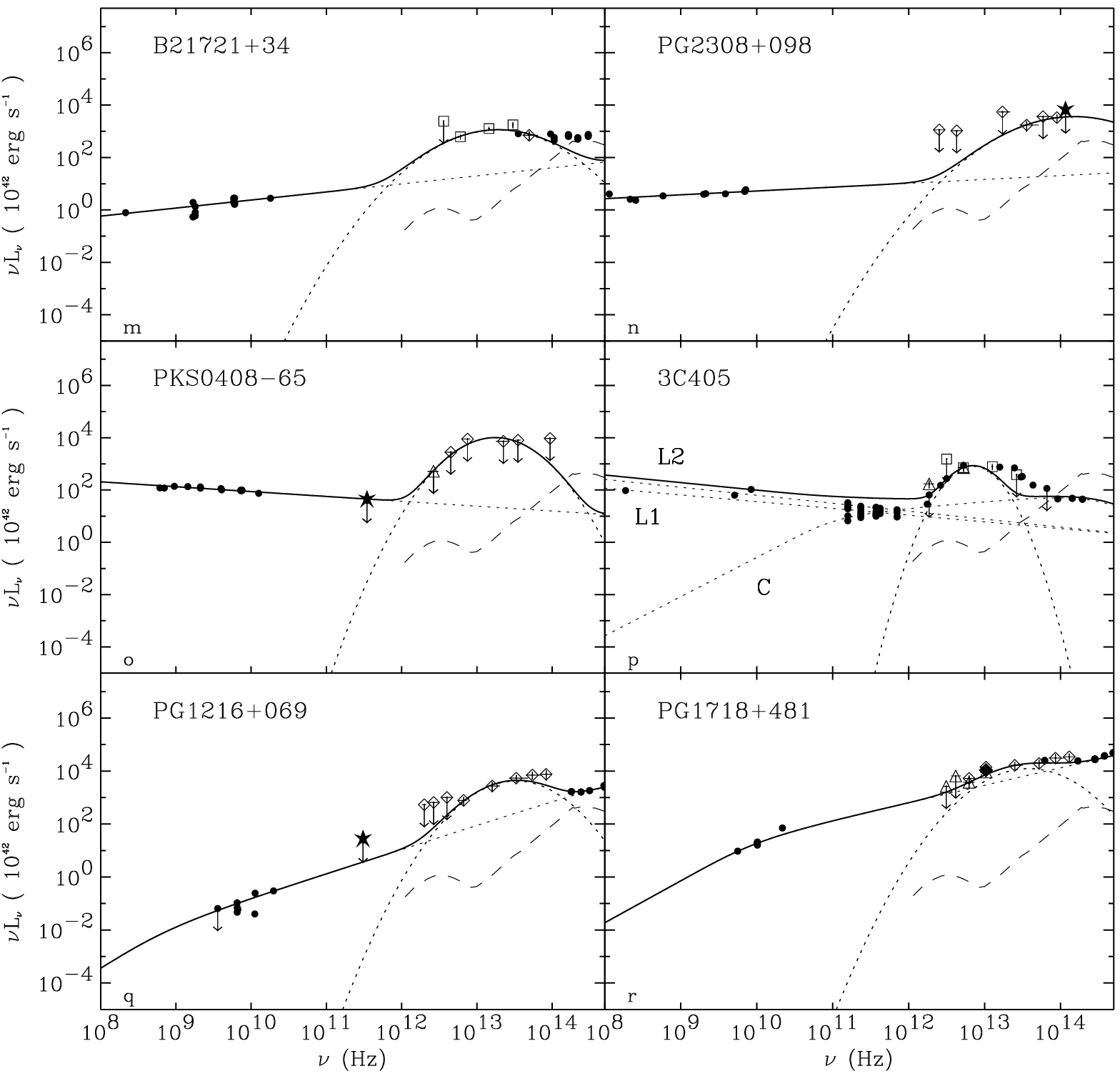}}
     \caption{Continued. The sources shown on this page include 2 SSRQ (m, n), 2 RG (o, p), and 2
      RIQ (q, r). In the case of PG 2308+098 the star represents the IRAC1 measurement.}
    \end{figure*}
%
   \begin{figure*}
      \setcounter{figure}{1}%
      \resizebox{14cm}{!}{\includegraphics{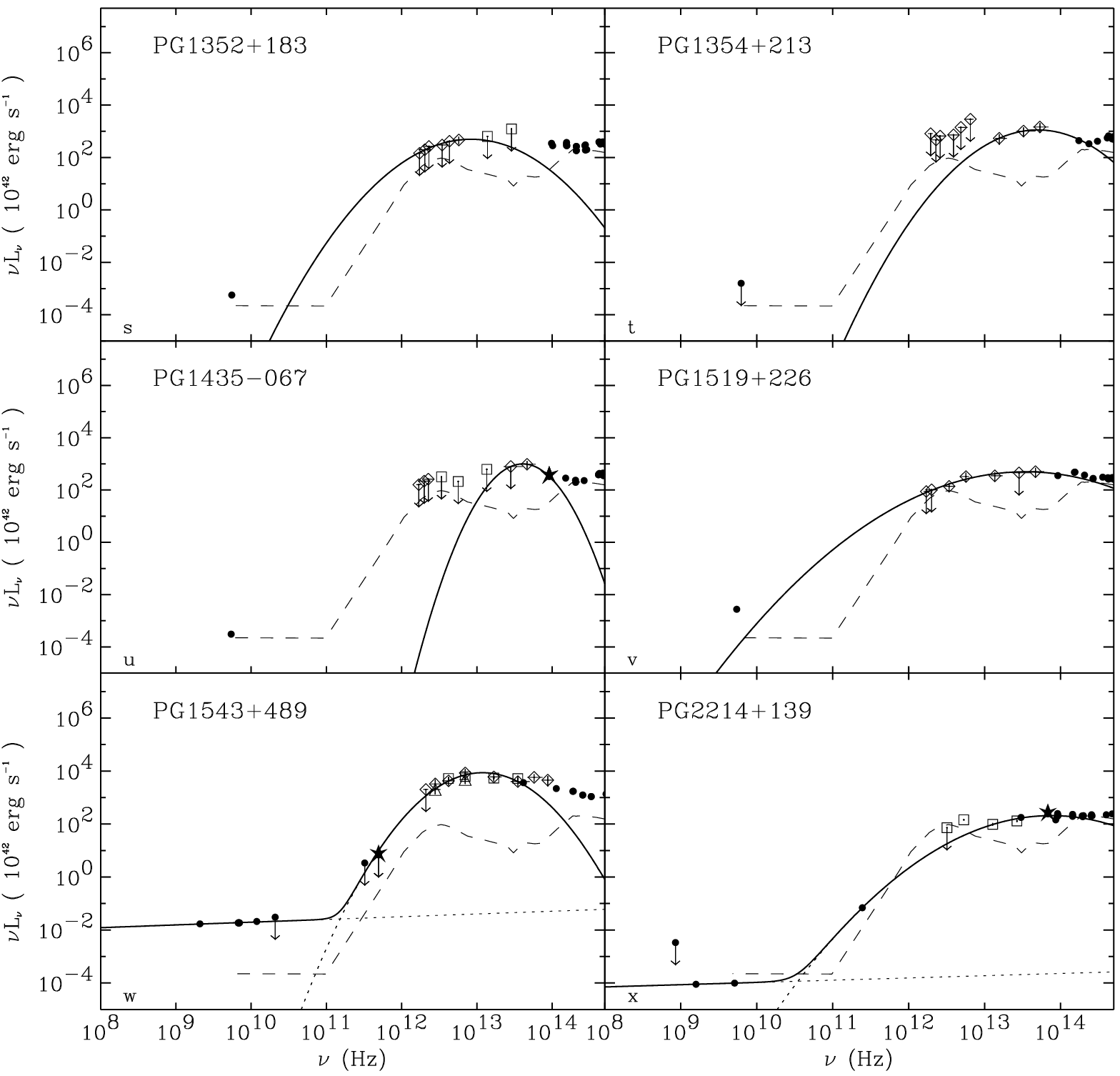}}
     \caption{Continued. All sources shown on this page are RQQ. In the case of
     PG 1435$-$067 the star represents the IRAC1 measurement.}
    \end{figure*}

\subsection{The nature of the IR emission: thermal or non-thermal emission ?}
\label{nature}
The IR emission can have a thermal or non-thermal origin. Several
investigation methods can be applied to identify the emission process:
\begin{enumerate}
\item The value of the slope of the sub-mm/far-IR spectral break can
      discriminate between optically-thick (self-absorbed) synchrotron and
      thermal emission from dust grains. The self-absorbed synchrotron model
      is characterized by a maximum value of the sub-mm/far-IR slope that is
      2.5 (F$_{\nu}\propto\nu^{\alpha}$), if the radiation is emitted by an
      electron population with a simple power-law energy distribution, or
      somewhat larger ($\alpha\leq$ 3) if a thermal electron pool or dual
      power-law energy distribution is invoked (\cite{deKool89};
      \cite{Schlickeiser91}). In either case the maximum synchrotron slope
      is attained only for a completely homogeneous source, otherwise it
      will be lower. The asymptotic thermal sub-mm/far-IR slope is
      expected to be $\geq$ 3 since the optically thin thermal
      spectrum derives from Rayleigh-Jeans law with an additional parameter
      dependent on frequency, $F_{\nu}\propto B_{\nu}(T)\cdot\tau_\mathrm{d}$
      where $\tau_\mathrm{d}$, the dust optical depth, is
      $\propto\nu^{\beta}$ with $\beta\simeq$1--2 (see section~\ref{Model_IR}).
\item A non-thermal origin of the IR emission is indicated if relatively
      short time scale flux variability is observed, since a synchrotron
      component is expected to come from a very compact source. Most of the
      dust emission comes from an extended source with a long variability
      time scale.
\item A thermal origin can be attributed to the IR emission if a non-varying
      excess from any reasonable extrapolation from the radio domain in the
      plot $\nu L_{\nu}$ versus $\nu$ is observed (\cite{Hughes97}).
\end{enumerate}
It is possible to distinguish the origin of the IR emission based
on brightness temperature (\cite{Sanders89}) and polarization
measurements, but the lack of these kind of data for the sample
does not allow us to apply them.

The first test will prove the thermal origin of the far-IR emission only if
the coldest dust component is also the brightest one. Colder and less bright
dust components will flatten the sub-mm/far-IR spectral slope. Two sources
in the sample, PG 1543+489 and 3C 405, have a sub-mm/far-IR spectral index
larger than 2.5. Their far-IR emission is therefore dominated by thermal
radiation. The remaining sources have insufficient data at long wavelengths
to constrain the sub-mm/far-IR spectral slope.

The observation of no variability is not conclusive, while if a flux
variation is observed the non-thermal hypothesis will be strongly supported.
This method can be applied only to those sources that were observed several
times. In our sample, at least two IR observations (IRAS and ISOPHOT, or
several ISOPHOT observations) at different epochs are available for ten
sources (3C 47, PKS 0637$-$752, PG 1100+772, PG 1543+489, PG 1718+481, B2
1721+481, HS 1946+7658, 3C 405, B2 2201+31A, and PG 2214+139) (see
Fig.~\ref{FigvarRLQ}). The data relative to the same observation epoch,
instrument and observation mode are represented with the same symbol and
connected by a line in Fig.~\ref{FigvarRLQ}. At least two measurements at
the same wavelength are available for all the ten objects with the exception
of PG 2214+139. Six sources (PG 1718+481, PG 1100+772, 3C 405, PG 1543+489,
B2 1721+34, and PG 2214+139) show no sign of variability. Two of them
also satisfy the first test. For two other sources (3C 47, and B2 2201+31A)
the observed variation is only marginally $<$1.6$\sigma$, where $\sigma$
includes the statistical and the systematic uncertainties. We consider
their emission as constant in the span of our observations. In the case
of PKS0637-752 we can only give a lower limit of the variation since the
source was not detected by ISOPHOT.  At wavelengths shorter than 60$\mu$m
ISOPHOT and IRAS give consistent results, while at longer wavelengths they
differ of more than 1.9$\sigma$.
  \begin{figure}
      \setcounter{figure}{1}%
      \resizebox{7.7cm}{!}{\includegraphics{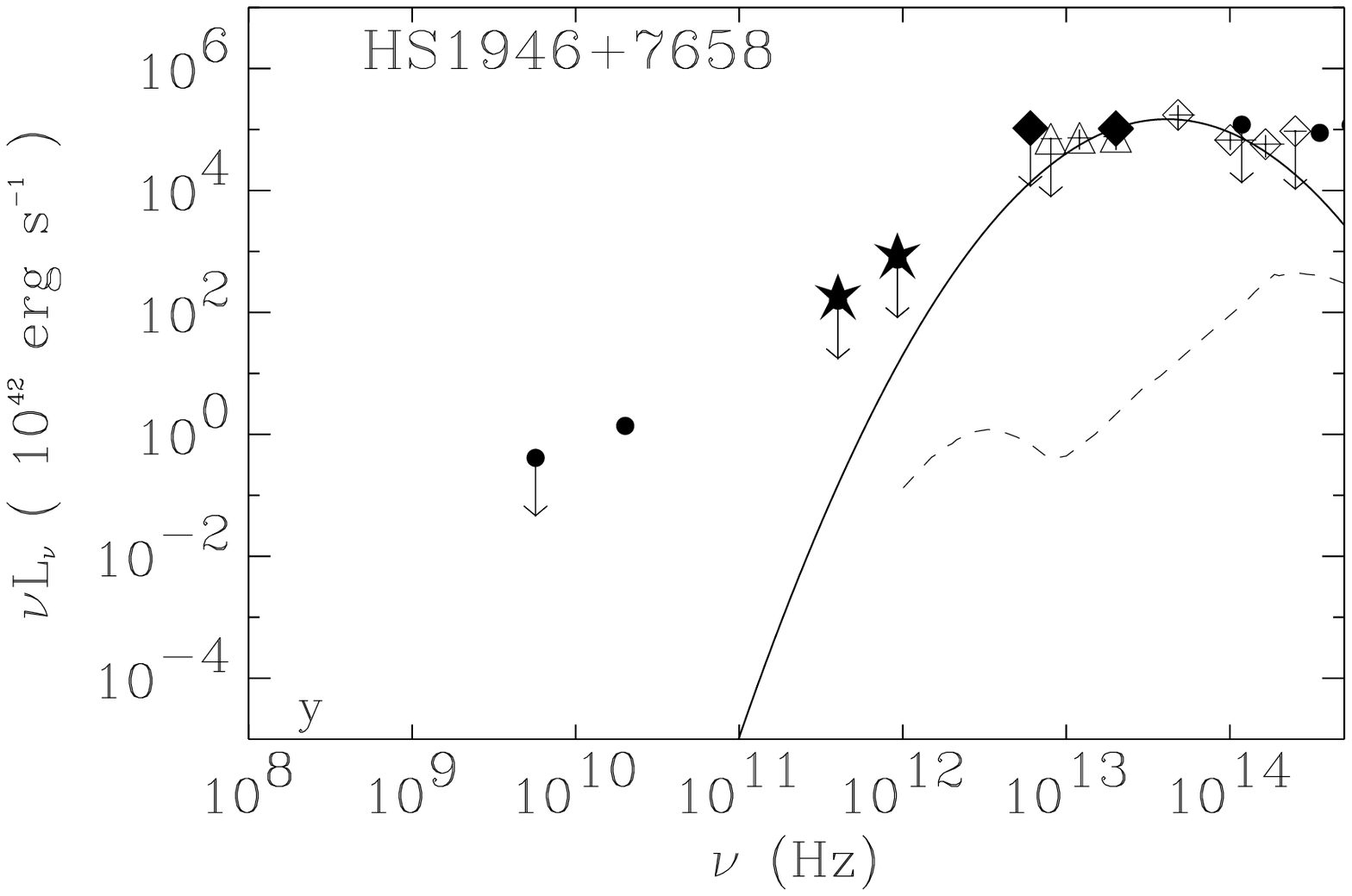}}
      \caption[]{Continued.}
         \label{Figsed1946+7658}
  \end{figure}
%

   \begin{figure*}
      \resizebox{14cm}{!}{\includegraphics{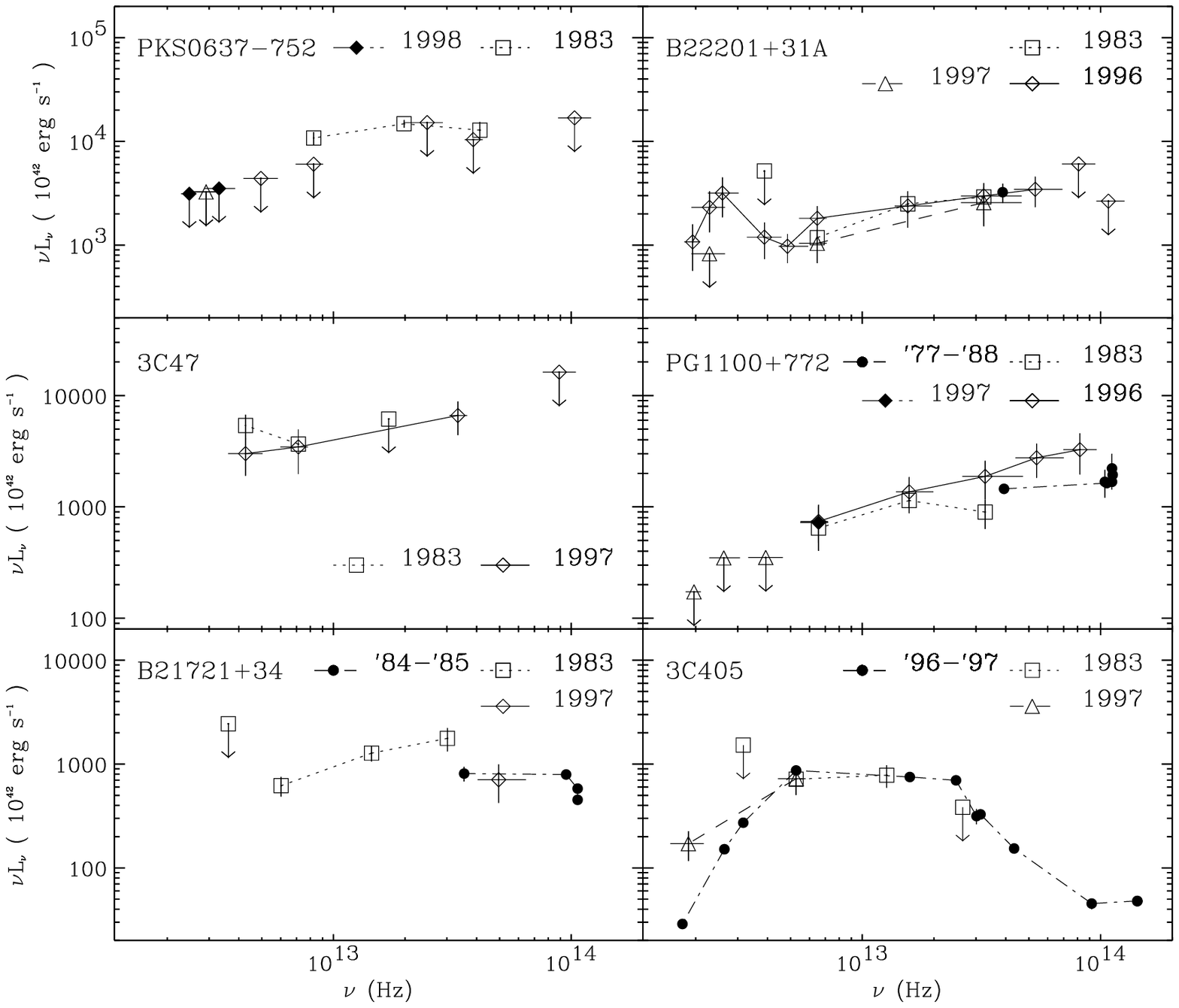}}
     \caption{IR data at different epochs. Symbols as in
     Fig.~\ref{FigsedFSRQ}. Lines connect data relative to the same
     epoch. All objects are RLQ (2 FSRQ, 3 SSRQ, and 1 RG).}
              \label{FigvarRLQ}
    \end{figure*}
   \begin{figure*}
      \setcounter{figure}{2}%
      \resizebox{14cm}{!}{\includegraphics{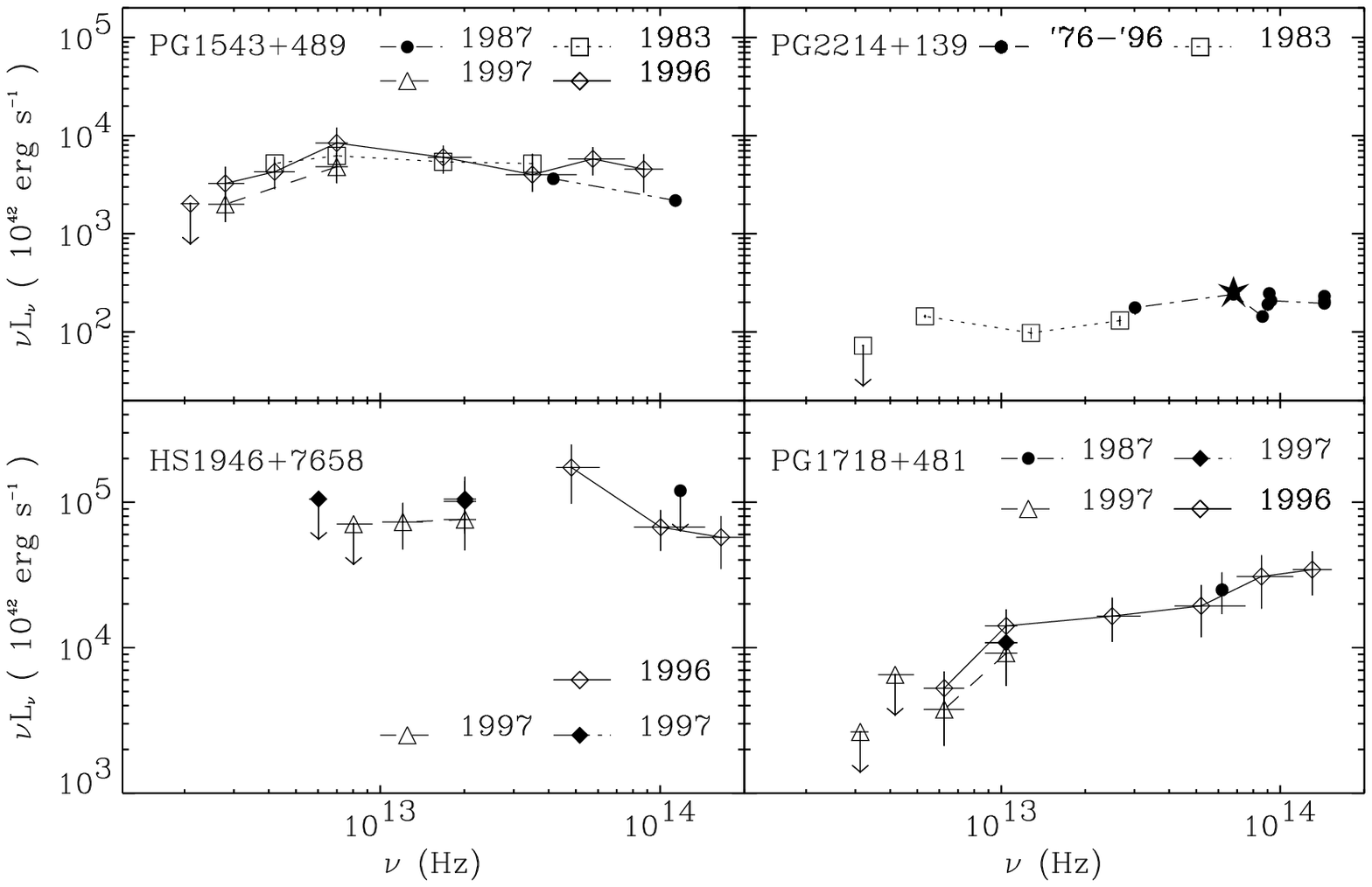}}
     \caption{Continued (3 RQQ, and 1 RIQ). }
    \end{figure*}
%
 \begin{figure}
 \centerline{ \resizebox{6.8cm}{!}{\includegraphics{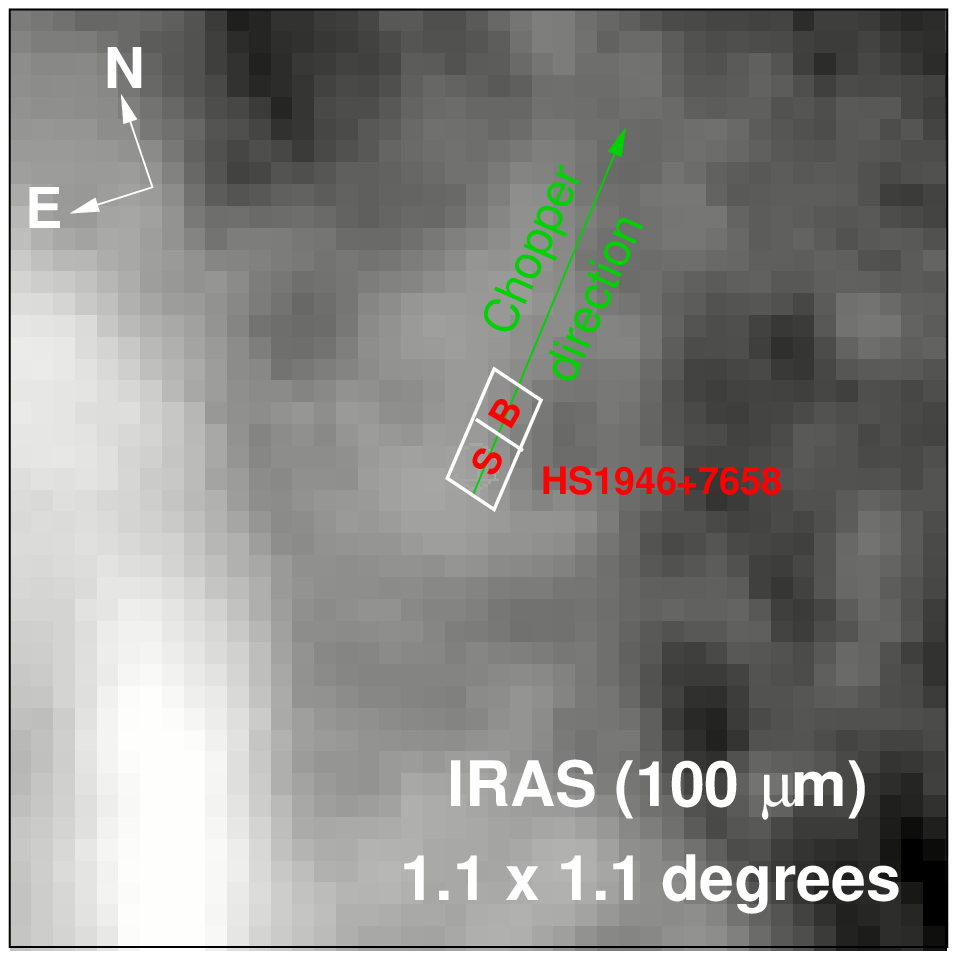}}}
     \caption[]{IRAS observation at 100 $\mu$m of a sky region of
                1.1$\times$1.1 degrees around the source HS 1946+7658, whose
                position is indicated by S. The sky region where the
                background was measured is indicated by B. The white polygon
                represents approximatively the region observed by the C200
                camera. The solid line indicates the chopper direction
                $\sim$41$^{\circ}$ to the West of North.}
        \label{Bg_100_HS1946}
 \end{figure}

The 150$\mu$m flux values for the high redshift source HS 1946+7658
differ by at least a factor of 6 between two epochs separated by less than a
year (see Tables~\ref{DetObsChop} and~\ref{DetObsRast}). Such extreme FIR
variability is unlikely in a RQQ. The larger flux resulted from a chopped
measurement, which is more susceptible at longer wavelengths to both
background fluctuations and instrumental effects (e.g., a number of
inadequately corrected cosmic ray strikes, unusually high detector drift,
etc...) than the raster maps. Fig.~\ref{Bg_100_HS1946} shows a 1.1$\times$1.1
degree region around HS 1946+7658 at 100$\mu$m from IRAS, with the position
of the source and the ISOPHOT chopper direction indicated. Bright cirrus
structures are nearby, and the background ranges from 0.21 -- 0.38 MJy/sr
within 90\arcsec of the source (equivalent to 0.16 -- 0.28 Jy on the C200
covered area) (IRSKY\footnote{IRSKY is An Observation Planning Tool for the
Infrared Sky developed for NASA at IPAC, JPL/Caltech.} version 2.5). A
strong gradient in the direction SE--NW in the background emission is
clearly shown, which coincides with the ISOPHOT chopper direction. These
variations may be greater at longer wavelengths and with the better spatial
resolution of ISOPHOT. As a consequence of the uncertainties for this
source, chopped data at $\lambda$ $>$ 100$\mu$m for HS 1946+7658 will not be
included in subsequent analysis.

The result of this method suggests that for all the selected objects,
with the exception of PKS 0637$-$752, the IR emission is thermal in origin.
The result is not very strong since a non-thermal process may also produce a
constant flux. However, it is unlikely to measure the same flux from a
non-thermal source during three different observations performed in a range
of 14 years (from 1983 to 1997) as observed in six objects (see
Fig.~\ref{FigvarRLQ}).

The third test is the simplest, and it can be applied to the whole sample
since a broad spectral coverage from radio to the IR is available for all
the sources. Before applying this test we need to estimate the
contribution from the non-thermal component in the IR and subtract it from
the observed IR spectrum. The non-thermal contribution can be estimated by
fitting the radio data with a reasonable model and then extrapolating it to
higher frequencies (see next section).

\subsection{Contribution of the radio non-thermal component in the infrared}
\label{rad_cont}

In order to estimate the contribution of the radio non-thermal component in
the IR domain, the radio continuum was fitted with some plausible models,
and extrapolated to higher frequencies. In the following we will distinguish
two components in the radio spectra of RLQ, the extended component (radio
lobes), and the core component (unresolved core, jet, etc...).

The radiation emitted by compact sources, like the cores and the hot-spots
of radio loud objects, can be modeled by a self-absorbed synchrotron emission
spectrum. In the case of a homogeneous plasma with isotropic pitch
angle distribution and power law energy distribution of the form
$N(E)\!\propto\!E^{-s}$, this can be expressed as:

\begin{equation}
L_{\nu}\propto\left(\frac{\nu}{\nu_\mathrm{t}}\right)^{\alpha_\mathrm{1}}\cdot
\left[1 - \exp \left(-\left(\frac{\nu_\mathrm{t}}{{\nu}}\right)^{\alpha_\mathrm{1}-\alpha_\mathrm{2}}\right)\right]
\cdot e^{-\frac{\nu}{\nu_\mathrm{cut\,off}}},
\label{sa_sync}
\end{equation}
where $\nu_\mathrm{t}$ is the frequency at which the optical depth of the plasma is
equal to unity, and $\nu_\mathrm{cut\,off}$ is the frequency corresponding to the
cut off energy of the plasma energy distribution, at which the
energy gains and losses of the electrons are equal. The optically thick and
optically thin spectral indices are denoted by $\alpha_\mathrm{1}$, and
$\alpha_\mathrm{2}$. In the case of a homogeneous source,
$\alpha_\mathrm{1}$ is expected to be 2.5, and $\alpha_\mathrm{2}$ is
related to the exponent $s$ of the plasma energy distribution by the
relationship:
$\alpha_\mathrm{2} = - (s - 1)/2$. The superposition of several self-absorbed
components can produce a flatter power law. In this case the spectral model of
equation~(\ref{sa_sync}) will remain valid, but $\alpha_\mathrm{2}$ will not have
the same meaning. 
If the source is optically thin, as in extended sources (radio lobes), the
emitted spectrum can be expressed more simply as:
\begin{equation}
 L_{\nu}\propto\left(\frac{\nu}{\nu_\mathrm{cut\,off}}\right)^{\alpha_\mathrm{2}}\cdot
\mathrm{e}^{-\frac{\nu}{\nu_\mathrm{cut\,off}}}.
\end{equation}
In many cases the contribution of the synchrotron component at high
frequencies is negligible compared to the observed emission,
therefore a high energy cut off was not included in the model.
The model is then represented by a broken power law of slopes
$\alpha_\mathrm{1}$ and $\alpha_\mathrm{2}$, or by a simple power law of
slope $\alpha_\mathrm{2}$, according to the presence/absence of self absorption.

The synchrotron model describes well the observed radio spectra of most of
the sources in the sample. However, while in the case of SSRQ it is quite
easy to separate the different components and hence apply a model for each
of them, the spectral modeling is more difficult for FSRQ. For these sources
we parameterize the emitted spectrum with an empirical equation that is
valid if the resulting spectrum is produced by self-absorbed synchrotron
emission or by optically thin synchrotron emission due to a hard electron
spectrum produced through the acceleration processes in turbulent plasma
(\cite{Wang97}). The observed flat radio spectra are described by the
following equation:
\begin{equation}
L_{\nu}\propto\left(\frac{\nu}{\nu_\mathrm{b}}\right)^{\alpha_\mathrm{1}}\cdot
\frac{\left[1 - \exp \left(-\left(\frac{\nu_\mathrm{t}}{{\nu}}\right)^{\alpha_\mathrm{1}-\alpha_\mathrm{2}}\right)\right]}
{\left[1 - \exp \left(-\left(\frac{\nu_\mathrm{b}}{{\nu}}\right)^{\alpha_\mathrm{0}-\alpha_\mathrm{1}}\right)\right]}
\cdot \mathrm{e}^{-\frac{\nu}{\nu_\mathrm{cut\,off}}},
\label{shock_model}
\end{equation}
where $\nu_\mathrm{b}$ is the frequency at which the spectrum flattens,
$\alpha_\mathrm{0}$ is the spectral index observed at low frequencies
($\nu<\nu_\mathrm{b}$), and the other model parameters have the same meaning
as in equation~(\ref{sa_sync}).

The observed SEDs from the radio to the mm energy domains were
fitted with one or a combination of these models (see best fit parameters in
Table~\ref{nonthermal_param}). 
   \begin{table*}
     \caption[]{Best fit parameters of non-thermal models}
     \label{nonthermal_param} 
      \begin{tabular}{@{}l c c  c c  r r@{}} 
      \hline 
\noalign{\smallskip}
\multicolumn{1}{c}{Source Name } & 
\multicolumn{1}{c}{Model$^{\dagger}$} &
\multicolumn{1}{c}{$\alpha_\mathrm{0}$} &
\multicolumn{1}{c}{$\alpha_\mathrm{1}$} &
\multicolumn{1}{c}{$\alpha_\mathrm{2}$} &
\multicolumn{1}{c}{$\nu_\mathrm{t}$} &
\multicolumn{1}{c}{$\nu_\mathrm{cut\,off}$} \\
\noalign{\smallskip}
\multicolumn{1}{c}{Source } & 
\multicolumn{1}{c}{} &
\multicolumn{1}{c}{} &
\multicolumn{1}{c}{} &
\multicolumn{1}{c}{} &
\multicolumn{1}{c}{(10$^{9}$ Hz)} &
\multicolumn{1}{c}{(10$^{14}$ Hz)} \\
\hline 
\noalign{\smallskip}
3C 47 (Core)          &  $b$ &        &  1.64  & $-$0.45 &   2.98   &	 \\
3C 47 (Lobe)          &  $c$ &        &	       & $-$0.99 &          &	 \\
PKS 0135$-$247        &  $d$  &$-$0.78 &  0.30  & $-$0.80 &  69.52   &  2.00 \\
PKS 0408$-$65         &  $c$ &        &	       & $-$1.19 &          &	 \\
PKS 0637$-$752        &  $d$  &$-$0.60 &  0.55  & $-$0.78 &  37.20   &  4.49 \\
PKS 0637$-$752 (Low)  &  $e$ &$-$0.55 &  0.55  & $-$1.50 &  75.12   &	 \\
PKS 0637$-$752 (High) &  $d$  &$-$0.60 &  0.49  & $-$0.78 &  41.28   &  4.22 \\
PG 1004+130 (Core)    &  $b$ &        &  0.53  & $-$0.19 &   2.00   &	 \\
PG 1004+130 (Lobe)    &  $c$ &        &        & $-$0.80 &          &	 \\
4C 61.20 (Core)       &  $b$ &        &  0.20  & $-$0.15 & 20.00 (F)&	 \\
4C 61.20 (Lobe 1)     &  $c$ &        &	       & $-$1.04 &          &	 \\
4C 61.20 (Lobe 2)     &  $c$ &        &	       & $-$0.93 &          &	 \\
PG 1048$-$090 (Core)  &  $b$ &        &  0.50  & $-$0.65 & 20.00 (F)&	 \\
PG 1048$-$090 (Lobe)  &  $c$ &        &	       & $-$0.84 &          &	 \\
PG 1100+772 (Core)    &  $b$ &        &  0.15  & $-$0.80 &  11.00   &	 \\
PG 1100+772 (Lobe)    &  $c$ &        &	       & $-$0.83 &          &	 \\
PG 1103$-$006 (Core)  &  $b$ &        &  0.30  & $-$0.50 & 20.00 (F)&	 \\
PG 1103$-$006 (Lobe)  &  $c$ &        &	       & $-$0.70 &          &	 \\
PG 1216+069           &  $a$  &        &  0.76  & $-$0.09 &   0.96   & 20.39 \\
PG 1543+489           &  $c$ &        &	       & $-$0.90 &          &	 \\
PG 1718+481           &  $b$ &        &  0.58  & $-$0.34 &  11.45   &	 \\
B2 1721+34            &  $c$ &        &	       & $-$0.69 &          &	 \\
3C 405 (Core)         &  $a$  &        &0.50 (F)& $-$0.74 &126.00 (F)&  3.58 \\
3C 405 (Lobe 1)       &  $c$ &        &	       & $-$1.26 &          &	 \\
3C 405 (Lobe 2)       &  $c$ &        &	       & $-$1.30 &          &	 \\
B2 2201+31A (Low)     &  $d$  &$-$0.52 &  0.60  & $-$0.30 &  28.60   &  0.034\\
B2 2201+31A           &  $d$  &$-$0.44 &  0.50  & $-$0.50 &  34.89   &  0.24 \\
PG 2214+139           &  $c$ &        &	       & $-$0.92 &          &	 \\
PG 2308+098           &  $c$ &        &        & $-$0.85 &          &       \\
\noalign{\smallskip}
\hline 
\end{tabular}\\
$^{\dagger}$ Model $a$ corresponds to equation~(\ref{sa_sync}), and model $b$ to the
same equation without cut off;\\$\,\,$ model $c$ corresponds to a simple power law; 
model $d$ to equation~(\ref{shock_model}), and model $e$ to the\\ 
same equation without cut off.\\
(F) indicates a fixed value.
\end{table*}
In some cases near-IR data have also been
used in the fits, in particular when no data in between mm and
near-IR frequencies constrained the spectrum to lie below the near-IR flux
(PKS 0135$-$247, PKS 0637$-$752, 4C 61.20, PG 1004+130, PG 1216+069, PG 1718+481,
3C 405, and B2 2201+31A). In a few cases we fixed some model parameters, as
indicated in a footnote of table~\ref{nonthermal_param}, since the available
data could not constrain them. The fixed values were chosen in the
range of values that provided reasonable spectra with properties similar to
those observed in other sources. Model $a$ in Table~\ref{nonthermal_param}
corresponds to equation~(\ref{sa_sync}). It was applied for modeling
core spectra of PG 1216+069, and 3C 405. The same model without the cut off ($b$)
was applied for modeling weak core spectra for which the high energy
cut off was not constrained (3C 47, PG 1004+130, 4C 61.20, PG 1048$-$090,
PG 1100+772, PG 1103$-$006, and PG 1718+481). A simple power law model ($c$)
was used to fit the radio emission from the lobes of SSRQ and RG (3C 47,
PKS 0408$-$65, PG 1004+130, 4C 61.20, PG 1048$-$090, PG 1100+772, PG 1103$-$006,
B2 1721+34, 3C 405 and PG 2308+098) and the radio emission of RQQ (PG 1543+489,
and PG 2214+139). Simultaneous observations available in the literature
generally do not provide wide or well-sampled wavelength coverage, so all
available data were used in the fits to the SEDs. The data and analysis are
adequate for the central purpose of estimating the contribution of the
non-thermal component to the IR emission. The model $d$, corresponding to
equation~(\ref{shock_model}), was applied to fit the radio emission of
FSRQ (PKS 0135$-$247, PKS 0637$-$752, and B2 2201+31A). The value of the
break frequency $\nu_\mathrm{b}$ was arbitrarily fixed to 2.75 GHz, since it
provides a good fit to the emitted spectrum of the three objects. In the
case of PKS 0637$-$752 (Low) (see section~\ref{unc_rad}) the cut off was not
included in the model ($e$) since the high frequency part of the spectrum is
very steep.

\subsubsection{Uncertainties in the radio contribution estimate}
\label{unc_rad}
The location of the high energy cut off is difficult to establish. Every
power law relative to the optically thin emission was extended at higher
frequencies until the spectrum turned down, and hence a cut off was required
by the data. A spectral cut off was thus required only in five objects (PKS
0135$-$247, PKS 0637$-$75, PG 1216+069, 3C 405, and B2 2201+31A), but it
could have been located at lower frequencies and present in other objects,
too. In most of the cases this parameter does not affect the presence and
the strength of the remaining IR flux, but its energy value may be important
in FSRQ (PKS 0135$-$247, PKS 0637$-$75, and B2 2201+31A), since these
objects have flat radio spectra for which extrapolation up to IR
frequencies is comparable to the IR fluxes. For these sources a more
accurate analysis of their radio spectra is needed. Since PKS 0135$-$247 was
not detected in the IR, no further analysis can be performed. We concentrate
only on PKS 0637$-$75 and B2 2201+31A. In order to better constrain the
non-thermal radio spectrum, i.e. to find some evidence of a spectral cut off
at sub-mm/far-IR frequencies, we searched in the literature for simultaneous
observations at these wavelengths, and we selected those that showed the
flattest and the steepest spectrum.
For PKS 0637$-$75 the flattest mm power law, chosen among several
simultaneous observations (\cite{Tornikoski96}), was measured on February
15th, 1990 ($\alpha$(3.0-1.3 mm) = $-$0.77), and the steepest one was
measured on April 4th, 1991 ($\alpha$(3.0-1.3 mm) = $-$1.47). The two power
laws are reported in Fig.~\ref{FigsedFSRQ}b with a dashed, and a
dashed-dotted line, respectively, plus displayed separately with flattest 
(Fig.~\ref{FigsedFSRQ}c) and steepest (Fig~\ref{FigsedFSRQ}d) spectral
fits.
The flattest spectrum overlaps the observed IR spectrum, leaving no
additional IR component. On the contrary, the extrapolation of the
steepest spectrum to IR frequencies is clearly below the observed IR
spectrum, but the IR observations were not simultaneous to the mm
observations. The source was observed by IRAS in 1983, and by ISO at
different wavelengths in 1997. During the elapsed time the source became
fainter in the far-IR, while shorter wavelength data from the two diferent
epochs are consistent. In the following we will suppose that a thermal IR
component is present, but dominating only at $\lambda <$60 $\mu$m, and we
will analyze its properties and compare them with those observed in other
sources.

For B2 2201+31A the flattest mm power law ($\alpha$(1.0-0.87 mm) =
$-$0.09) was measured on February 1989 (\cite{Chini89a}), and the
steepest one was measured on September 14th, 1993 ($\alpha$(2.0-1.3-1.1 mm)
= $-$0.72). The two power laws are reported in Fig.~\ref{FigsedFSRQ}e with a
dashed, and a dashed-dotted line, respectively. The spectrum is in both
cases quite flat, however the extrapolation of the 1993 spectrum lies below
the IR spectrum. More than the sub-mm data, the analysis of the emission at
shorter wavelengths gives important indications on the origin of the IR
emission. B2 2201+31A was observed on September 15th, 1993 also in the
near-IR (simultaneous sub-mm and near-IR data are indicated by large open
circles in Fig.~\ref{FigsedFSRQ}e). The near-IR data are above the
extrapolation of the sub-mm data, suggesting the presence of two different
spectral components in these two wavelength ranges (see the analogous case
of 3C 273 in Robson et~al. (1986)). This hypothesis is also suggested by the
constant emission observed up to 60 $\mu$m. A non-thermal source is expected
to vary more at higher frequencies, due to greater energy losses. All these
considerations suggest the short wavelength continuum is dominated by a
thermal component. As in the case of PKS 0637$-$75 we will suppose that an
additional IR thermal component is present at
$\lambda\,\leq$ 60 $\mu$m.

These two sources (PKS 0637$-$752, and B2 2201+31A) are good examples of how
variability can create an artificial IR spectral turnover, or hide a real
one. An IR spectral turnover may be due to different luminosity states of
the source at different epochs, instead of to the presence of a separate IR
component. The weakness of the radio emission in RQQ precludes that its
extrapolation could account for the IR emission for all reasonable
assumptions on the radio variability. In SSRQ the extrapolation of the radio
component in the IR is usually too faint to explain the IR emission, even if
we take into account variability. The variability factors observed in two
SSRQ in our sample, 3C 47 and PG 1004+130, are too small to explain the much
higher IR fluxes, and this is probably true for the SSRQ in general. In the
mm domain we measured a flux variation from the core of the SSRQ 3C 47 of a
factor of $\sim$ 2 in almost three years (the emitted flux density at $\sim$
100 GHz was equal to 16.3$\pm$0.9 mJy on September 1995
(\cite{vanBemmel98}), and equal to 30.8$\pm$0.6 mJy on July 1998 (this
work)). The SSRQ PG 1004+130 was observed twice at 6 cm, in 1982 and in
1984, with a flux variation of a factor $\sim$2.5, from 12 mJy to 30 mJy
(\cite{Lister94}).

In conclusion, the radio models shown in Fig.~\ref{FigsedFSRQ} indicate
the presence of an additional IR component in almost the whole sample.
According to the third test, this result indicates that the
observed IR emission is of thermal origin. The properties of the IR emission
in quasars will be derived and analyzed in section~\ref{Model_IR}, after
subtraction of the non-thermal contribution extrapolated from the radio
domain.

\subsection{Modeling of the IR component}
\label{Model_IR}

The IR emission can be accounted for by reradiation of the central
luminosity by gas and dust in warped discs in the host galaxies of the
quasars (\cite{Sanders89}), in the outer edge of the accretion disc and in a
torus of molecular gas within a few parsecs of the central energy source
(\cite{Niemeyer93}; \cite{Granato94}; \cite{Granato97}; \cite{Pier92},
1993), and/or by starburst emission (\cite{Rowan95}).
The host galaxy starlight contribution is probably negligible in the
far/mid-IR since the host galaxy spectrum largely differs in shape and
luminosity from the SED of the selected objects (see Fig.~\ref{FigsedFSRQ}).
We describe here the main observational properties of the different objects
of each class and compare them using a very simple model of thermal emission:
the grey body model. This model does not take into account the source
geometry (toroidal, warped disc, etc). An isothermal grey body at the
temperature T emits at frequency $\nu$ a luminosity density given by the following
equation~(\cite{Gear88},~\cite{Weedman86}):
\begin{equation}
   L_{\nu\,\mathrm{em}} = 4\pi^2r^2 \cdot \mathrm{B}(\nu_\mathrm{em},\mathrm{T}) \cdot (1-e^{-\tau_\mathrm{d}}), 
\end{equation}
where $r$ is the radius of the projected source, B($\nu$,T) is the Planck
function for a blackbody of temperature T, and $\tau_\mathrm{d}$ is the
optical depth of the dust. The optical depth can be approximated by a power
law of type $\tau_\mathrm{d}$ = $(\nu/\nu_\mathrm{0})^{\beta}$, where
$\nu_{\mathrm{0}}$ is the frequency at which the dust becomes optically
thin, and $\beta$ is the dust emissivity index. A non-linear least squares
fit was used in the fitting procedure, leaving the radius $r$, the
temperature T, and the frequency $\nu_\mathrm{0}$ free to vary , while the
emissivity exponent $\beta$ was fixed equal to 1.87 (\cite{Polletta99}).
   \begin{figure*}
      \resizebox{14cm}{!}{\includegraphics{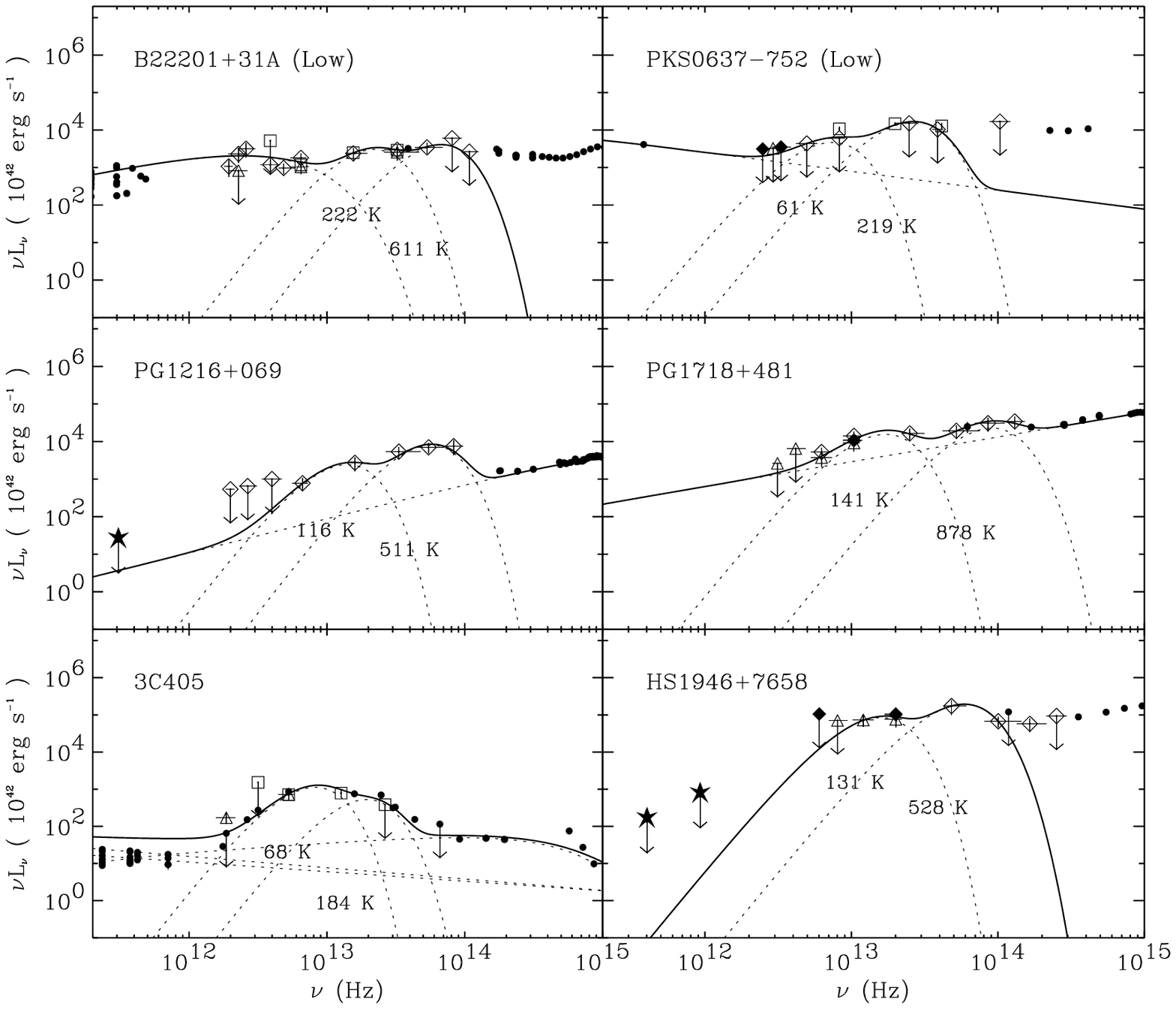}}
     \caption{SEDs as $\nu L_{\nu}$ versus $\nu$ in the rest frame of the
              objects (2 FSRQ, 2 RIQ, 1 RG, 1 RQQ). Symbols as in
              Fig.~\ref{FigsedFSRQ}. Dotted lines represent the best fit
              non-thermal models of the radio component and the best fit
              grey body models of the IR component.  The temperature of each
              grey body component is reported. The sum of all single fitted
              models is shown by a solid line. }
              \label{FiggbFSRQ}
    \end{figure*}

   \begin{figure*}
      \setcounter{figure}{4}%
      \resizebox{14cm}{!}{\includegraphics{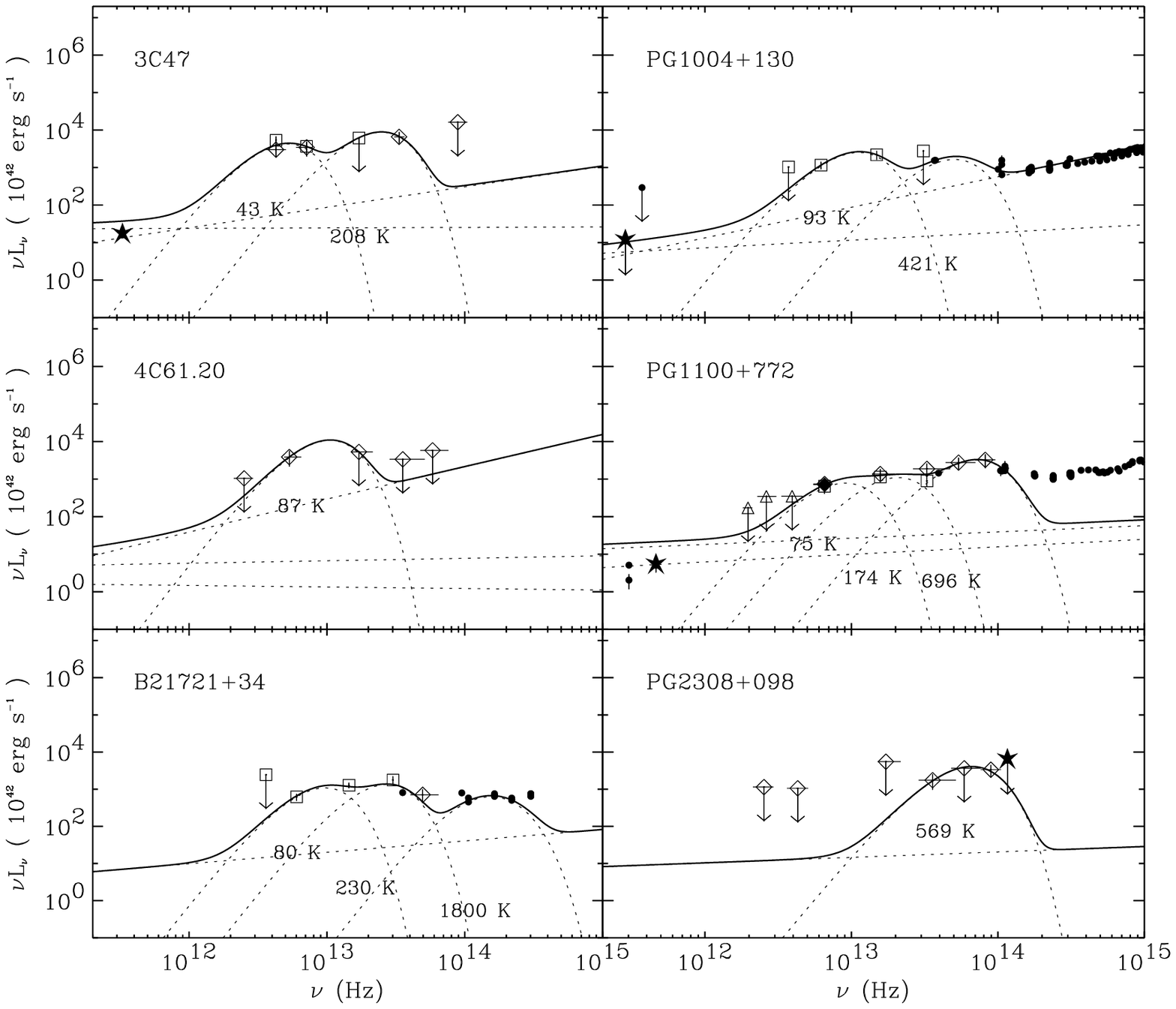}}
     \caption{Continued (6 SSRQ).}
    \end{figure*}
   \begin{figure*}
      \setcounter{figure}{4}%
      \resizebox{14cm}{!}{\includegraphics{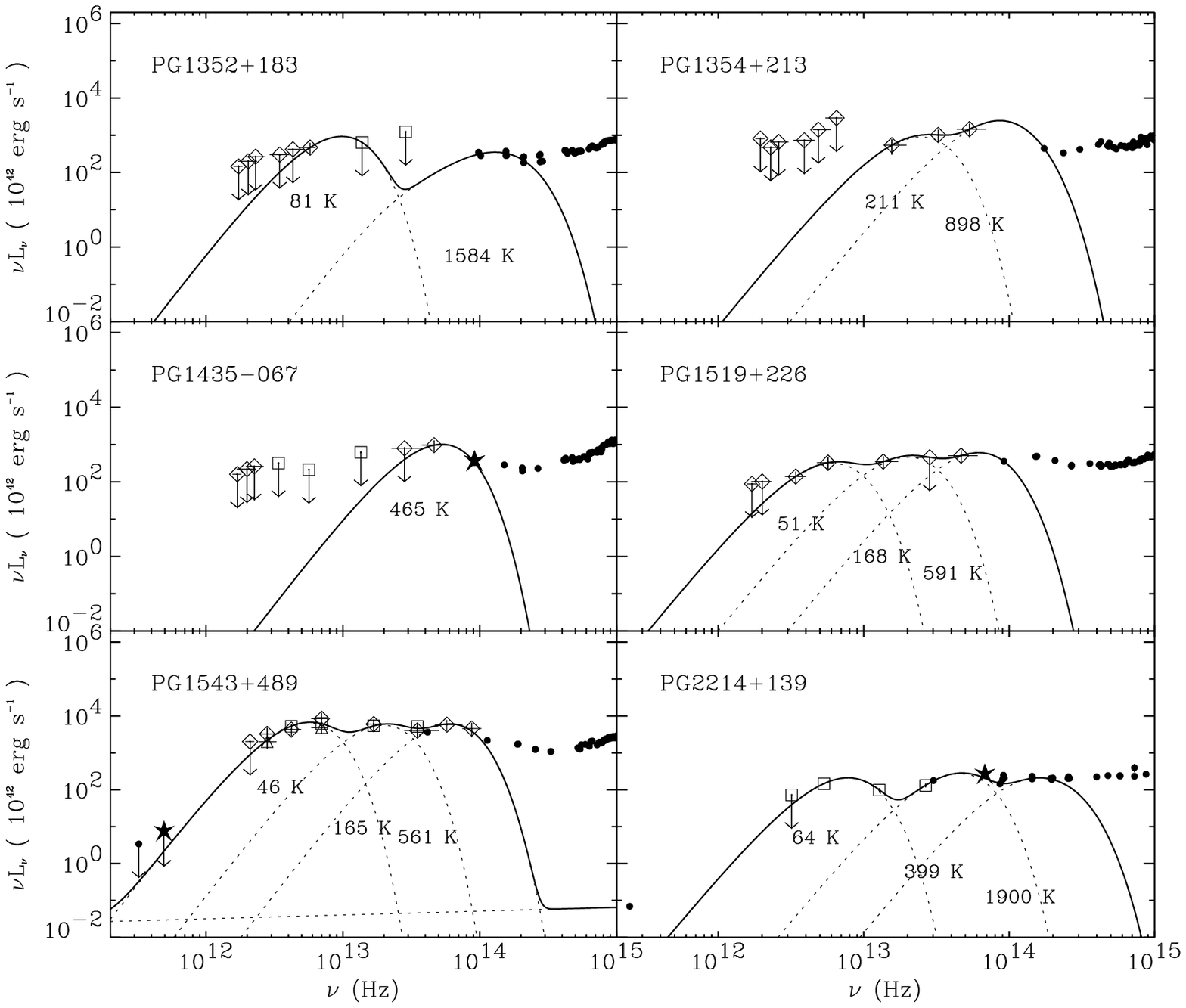}}
     \caption{Continued (6 RQQ).}
    \end{figure*}

\begin{figure}
      \resizebox{8.8cm}{!}{\includegraphics{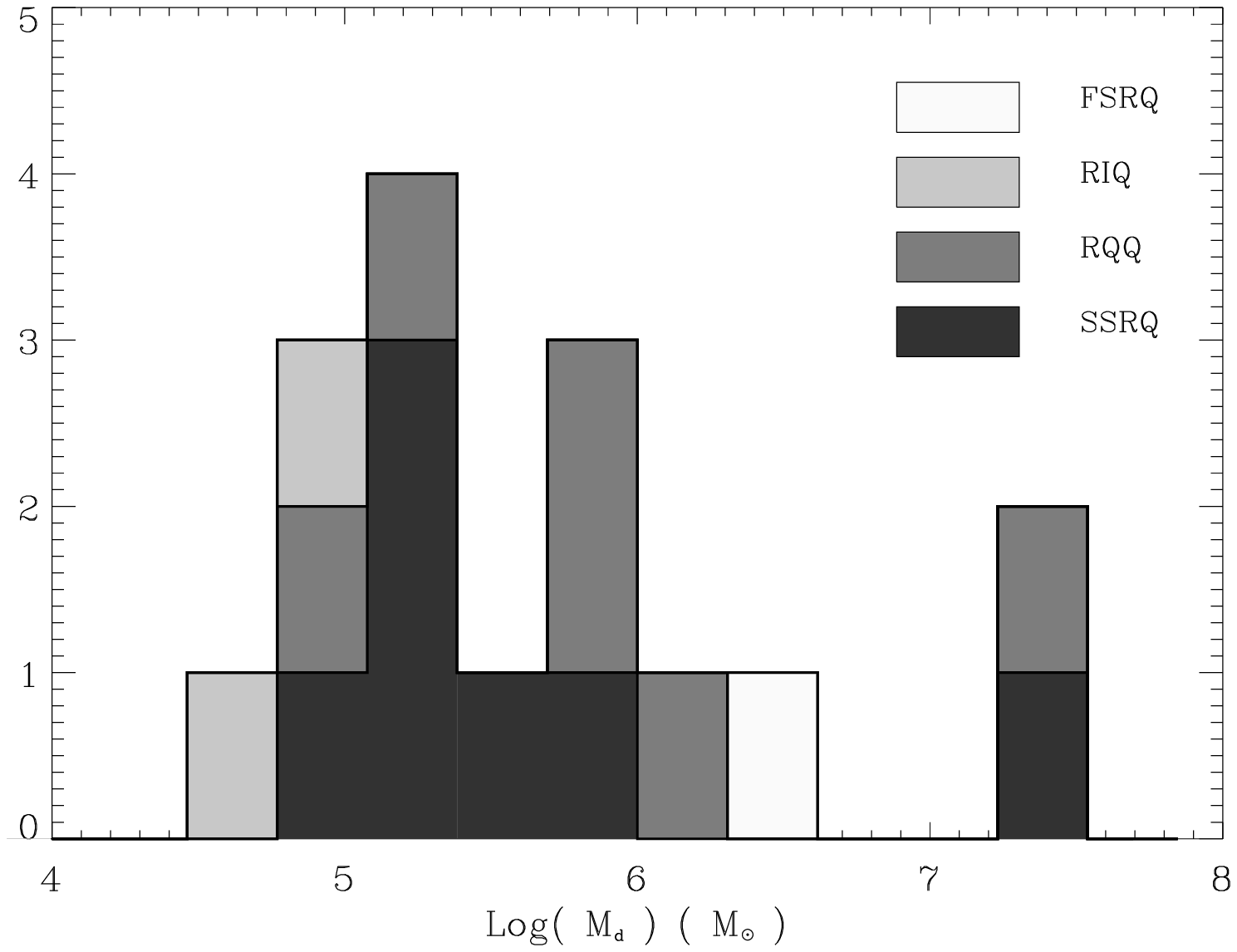}}
      \caption[]{Histogram of the dust masses for the different classes:
         RQQ, SSRQ, FSRQ, and RIQ.}
         \label{hist_mass}
  \end{figure}

The observed IR SEDs are smooth and indicate a wide and probably continuous
range of dust temperatures, describable by several grey body components. The
best fit grey body models of the observed IR SEDs are shown in
Fig.~\ref{FiggbFSRQ}. The thick solid line represents the sum of non-thermal
and grey body components. Each individual component is represented by a
dotted line.
The temperature (T) and the size (r) of each grey body component are listed
in columns 5--10 of Table~\ref{temp_size_param}. It is worth noting that we
could fit the observed IR spectra using a different optical depth function
(different $\beta$, and $\nu_\mathrm{0}$ values).
The optical depth value is important in a discussion of the source geometry
in terms of an extended or compact heating source. In our models the optical
depth values derived by the fits are low ($<\,$1) in the far/mid-IR, and
$\simeq$1 in the near-IR ($\sim$ 3$\mu$m). If the dust becomes optically
thin at longer wavelengths, the real source sizes will be smaller than our
estimates, and vice versa. Using our optical depth values, the estimated
sizes of the observed dust components range between 0.06\,pc and
9.0\,kpc, and the temperatures between 43\,K and 1900\,K.
The minimum temperature may be due to an absence of dust at large distances
(few kpcs) or at low temperature, and/or to starlight heating to the orders
of the inferred minima. The maximal temperature is generally explained as a
drop in opacity caused by the sublimation of the most refractory grains at
temperatures T $\sim$ 2000 K (\cite{Sanders89}). The total luminosities
observed in the IR, obtained by integrating the grey body components (see
column 1 in Table~\ref{temp_size_param}), vary over a wide range, from
2.0$\cdot$10$^{11}$ $L_\mathrm{\odot}$ to 7.6$\cdot$10$^{13}$
$L_\mathrm{\odot}$. No significant difference in the distribution of sizes,
temperatures, and luminosities are observed among different types of
quasars.
   \begin{table*}
     \caption[]{Best fit parameters of grey body models}
     \label{temp_size_param} 
      \begin{tabular}{@{}l@{}c@{}c@{}c@{}c@{}r@{}c@{}c@{}r@{}c@{}} 
      \hline 
\noalign{\smallskip}
\multicolumn{1}{c}{Source Name} & 
\multicolumn{1}{c}{Total L(IR)} &
\multicolumn{1}{c}{$\frac{L(60-200\mu m)}{L(3-60\mu m)}$} &
\multicolumn{1}{c}{Log(M$_\mathrm{d}$)} &
\multicolumn{2}{c}{I Component} &
\multicolumn{2}{c}{II Component} &
\multicolumn{2}{c}{III Component} \\
\multicolumn{1}{c}{ } & 
\multicolumn{1}{c}{(10$^{11}$ L$_\mathrm{\odot}$)} &
\multicolumn{1}{c}{ } & 
\multicolumn{1}{c}{(M$_\mathrm{\odot}$)} &
\multicolumn{1}{c}{T (K)} &
\multicolumn{1}{c}{r (pc)} &
\multicolumn{1}{c}{T (K)} &
\multicolumn{1}{c}{r (pc)} &
\multicolumn{1}{c}{T (K)} &
\multicolumn{1}{c}{r (pc)} \\
\hline 
\noalign{\smallskip}
3C 47                &  36.1  &  0.205  &   7.55  &  42.7 &  7621 &  208 & 109 &       &       \\
PKS 0637$-$752 (Low) &  58.6  &  0.056  &   6.68  &  61.0 &  1923 &  219 & 276 &       &       \\
PG 1004+130          &  11.6  &  0.040  &   5.35  &  92.6 &   663 &  421 &  12 &       &       \\
4C 61.20             &  29.8  &  0.086  &   6.14  &  86.8 &  1401 &      &     &       &       \\
PG 1100+772          &  14.1  &  0.024  &   5.38  &  74.9 &   597 &  174 &  49 &   696 &  2.16 \\
PG 1103$-$006        &  25.3  &  0.024  &   5.18  & 120.0 &   496 &      &     &  1322 &  0.33 \\
PG 1216+069          &  28.8  &  0.007  &   4.76  & 115.8 &   310 &  511 &   8 &       &       \\
PG 1352+183          &   3.8  &  0.091  &   5.26  &  80.9 &   556 &      &     &  1584 &  0.11 \\
PG 1354+213          &   9.3  &  0.001  &         &       &       &  211 &  23 &   898 &  0.98 \\
PG 1435$-$067        &   2.8  &         &         &       &       &      &   4 &   465 &       \\
PG 1519+226          &   3.8  &  0.097  &   5.97  &  50.7 &  1264 &  168 &  41 &   591 &  1.34 \\
PG 1543+489          &  49.9  &  0.202  &   7.54  &  46.0 &  8997 &  165 & 183 &   561 &  5.17 \\
PG 1718+481          & 105.3  &  0.007  &   5.06  & 141.5 &   573 &      &     &   878 &  3.60 \\
B2 1721+34           &   8.6  &  0.047  &   5.36  &  80.0 &   611 &  230 &  32 &  1800 &  0.11 \\
HS 1946+7658         & 761.5  &  0.006  &   6.00  & 130.8 &  1294 &  528 &  37 &       &       \\
3C 405               &   4.6  &  0.137  &   5.79  &  67.7 &  1265 &  184 &  13 &       &       \\
B2 2201+31A (Low)    &  19.5  &  0.002  &         &       &       &  222 &  18 &   611 &  3.99 \\
PG 2214+139          &   2.0  &  0.091  &   5.22  &  63.7 &   540 &  399 &   3 &  1900 &  0.06 \\
PG 2308+098          &  11.1  &         &         &       &       &      &   6 &  569  &       \\
\noalign{\smallskip}
\hline 
\end{tabular}
\end{table*}
%
We also derive the mass of each dust component at the measured temperature,
using the following equation (\cite{Hughes97}):
\begin{equation}
 \mathrm{M}_\mathrm{d} =
\frac{L_{\nu\,\mathrm{em}}}{K_\mathrm{d}(\nu_\mathrm{emis})}\frac{1}{\mathrm{B}(\nu_\mathrm{emis},\mathrm{T})},
\end{equation}
where $K_\mathrm{d}(\nu_\mathrm{emis}) \propto \nu^{\beta}$, $\beta$ = 1.87,
is the rest-frequency dust absorption coefficient. The normalization is
$K_{\rm d}(\nu_{\rm emis}) = 10\; {\rm cm}^2 {\rm g}^{-1}$ at 250
$\mu$m (\cite{Hildebrand83}), giving $K_\mathrm{d}(\nu)$ = 1.14 cm$^{2}$
g$^{-1}$ at 800 $\mu$m. The range of assumed values of
$K_\mathrm{d}(\nu)$ at 800 $\mu$m in the literature is 0.4--3.0 cm$^{2}$
g$^{-1}$ (\cite{Draine84};~\cite{Mathis89}). Our dust mass estimates can
thus differ by, at most, a factor 2.7.
The derived values of dust masses are reported in column 4 of
Table~\ref{temp_size_param}, and, separately for each class, in
Fig.~\ref{hist_mass}. 
Since the largest dust masses are located in the outer, less illuminated,
lower temperature regions of the dust distribution, M$_\mathrm{d}$ is mainly
constrained by far-IR data. Therefore, when sub-mm and far-IR data are not
available, the real dust mass cannot be well measured. For this reason we
did not report the dust and gas masses when the low temperature component
was not constrained. The absence of data in the near-IR has a negligible
effect on the dust mass estimate. As for the other parameters (T, L(IR), r),
the dust mass distribution does not differ significantly among different
types of quasars (see Fig.~\ref{hist_mass}).

\section{Similarities and differences in the SED of RLQ and RQQ}

The present sample contains a range of radio source classifications, with
which we can elucidate the dependence of broad-band spectral features on
radio properties, thereby testing some unification scenario predictions.
The limitations of these tests lie in the sample's relatively small size
and heterogeneous nature.

\subsection{Average SED}
\label{avgsedpar}

\begin{figure*}
      \resizebox{12cm}{!}{\includegraphics{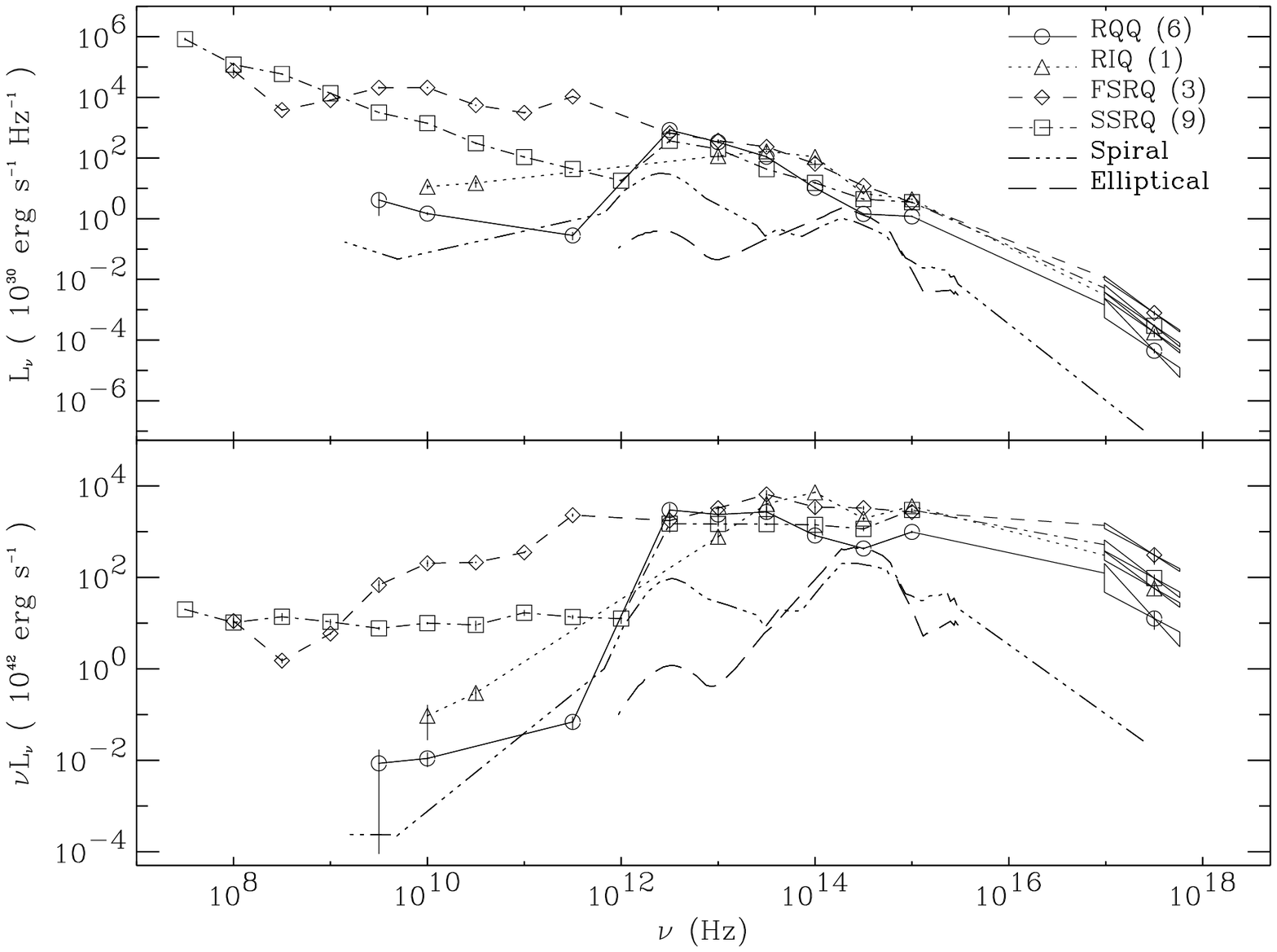}}
      \hfill\parbox[b]{5.5cm}{\caption[]{The average quasar energy
		distribution for RQQ (solid line), RIQ (dotted line), FSRQ
		(dashed line), and SSRQ (dash-dot line). The number of
		objects that were used in each class is reported 
		on the figure with the class name. A symbol is
		reported for each frequency bin, if at least one data point is
		available. The two high redshift objects of the sample:
		HS 1946+7658, and PG 1718+481, were excluded since they
		are characterized by very high luminosities that modify the
		average SEDs at a few frequencies producing an irregular spectral
		shape. Two host galaxy templates are also reported: a
		typical spiral galaxy (dash-dot-dot-dot line), and a giant
		elliptical galaxy (long dashed line). No normalization
		was applied to the reported curves.}
         \label{avg_SED}}
  \end{figure*}

A quick look at the main spectral differences between the different kinds of
quasars is provided by the comparison of the average SED of RQQ, RIQ, FSRQ,
and SSRQ (including the RG 3C 405). The SEDs are shown in $L_{\nu}$--$\nu$
and $\nu$L$_{\nu}$--$\nu$ spaces separately for each class over the
radio/soft X-ray frequency range in Fig.~\ref{avg_SED}.
The broad spectra of two typical host galaxies (a giant elliptical and a
spiral galaxy), in their rest frames and without any normalization, are
also plotted in Fig.~\ref{avg_SED}, as in Fig.~\ref{FigsedFSRQ}.
The average SEDs have been computed using the conventional mean, excluding
upper limits. The width of each frequency bin is equal to 0.5 in
Log($\nu$). The reported uncertainties correspond to the standard deviation
of the mean of the data per frequency bin. All the data have been
connected by straight lines. At soft X-ray energies we
indicate the average power law $\pm$ 1$\sigma$
computed from the distribution of best fit soft X-ray power law models of
all objects of the same class (photon index $\Gamma$ = 2.73$\pm$0.61 (RQQ),
2.39$\pm$0.19 (RIQ), 2.39$\pm$0.23 (SSRQ), and 2.25$\pm$0.12 (FSRQ)).

As expected, the largest difference in luminosity among the different
classes appears at radio wavelengths. A smaller difference is observed at
soft X-ray energies, and in the near-IR ($\nu >$ 10$^{14}$ Hz corresponding
to $\lambda <$ 3 $\mu$m), while the luminosity and the spectral shape in the
mid- and far-IR are remarkably similar (see Fig.\ref{avg_SED}). The large
difference in the IR spectral shape between quasars and the host galaxy
templates indicates that the contribution from the host galaxy is negligible
also at radio and soft X-ray energies, and not only in the far/mid-IR.
This result is in agreement with previous studies on the broad SED of
quasars (\cite{Sanders89};~\cite{Elvis94}). A quantitative comparison of the
luminosity emitted at different frequencies by each quasar class is
presented in the next section.

\subsection{Multi band luminosities}
\label{multibands}

\begin{figure}
      \resizebox{8.8cm}{!}{\includegraphics{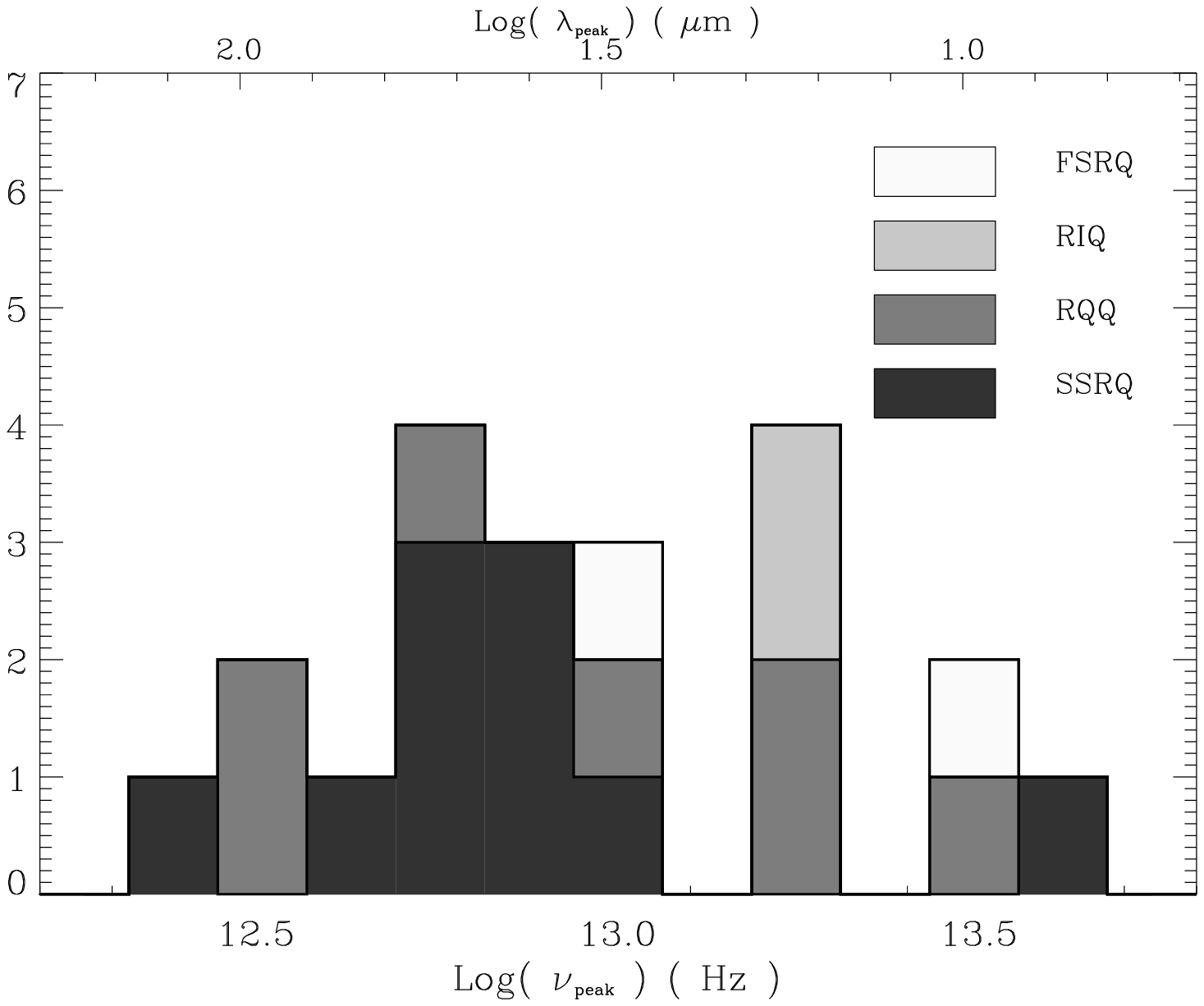}}
      \caption[]{Histogram of the IR peak frequency of the parabola
                 model for the different classes: RQQ,
                 SSRQ (including the RG 3C 405), FSRQ, and RIQ.}
         \label{hist_numax}
  \end{figure}

The IR component was also modeled by fitting a parabola in Log
L$_{\nu}$--Log $\nu$ space (see Fig.~\ref{FigsedFSRQ}). The parabola
model gives a rough estimate of the strength and shape of the IR component,
even if the spectral coverage is not complete. For several objects upper
limits were also used in the fit. This model has by itself no physical
meaning, however, it describes the IR component relatively well, it can
easily be traced even with poor spectral coverage, and can take into account
the whole IR emission of most of the objects in a larger wavelength range
than the detailed grey body models. The parabola is too narrow to satisfy
the observed IR SED in a few cases, e.g., in 3C 405 and PG 1543+489. In
these cases we fitted only the far/mid-IR data where the IR emission usually
peaks. The parabola parameters are its width, the frequency of maximum
luminosity density ($\nu_\mathrm{peak}$) and the maximum luminosity density
($L_{\nu_\mathrm{peak}}$).  The parabola fit to the IR component was applied
to all objects of the sample, except PKS 0135$-$247, for which no IR data
are available (see Fig.~\ref{FigsedFSRQ}a). The distribution of the peak
frequencies values ($\nu_\mathrm{peak}$) observed in the four different
classes of objects is reported in Fig.~\ref{hist_numax}. The distribution is
quite similar for SSRQ and RQQ, ranging from 2.6$\cdot$10$^{12}$ Hz (114
$\mu$m) to 3.6$\cdot$10$^{13}$ Hz (8 $\mu$m), while it is shifted to higher
frequencies for FSRQ and RIQ, ranging from 9.0$\cdot$10$^{12}$ Hz (33
$\mu$m) to 2.8$\cdot$10$^{13}$ Hz (11 $\mu$m). This difference may be due to
the flat radio non-thermal component extending to high frequencies in FSRQ
and RIQ, and dominating the dust emission.

\begin{figure}
      \resizebox{8.8cm}{!}{\includegraphics{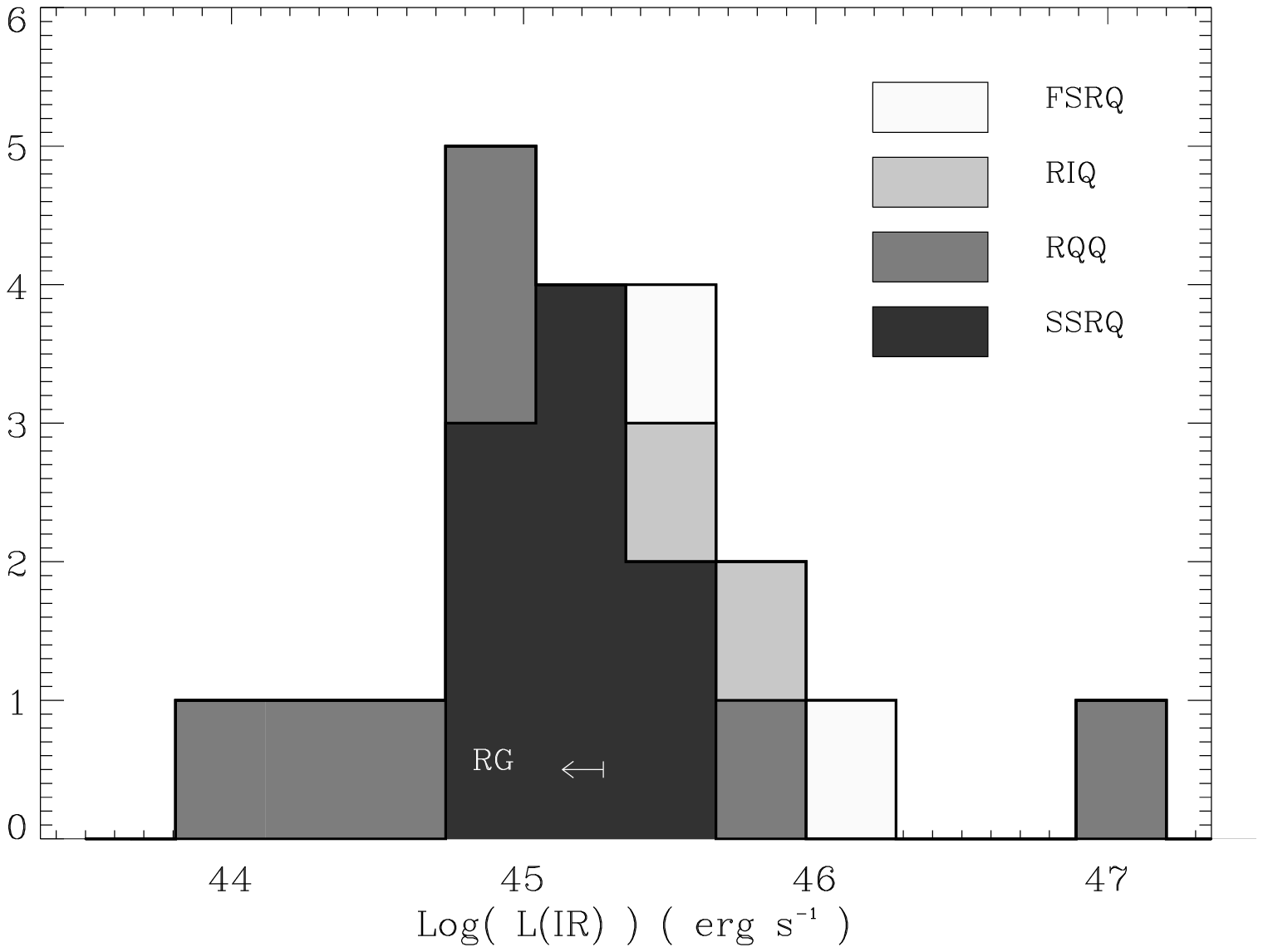}}
      \caption[]{Histogram of the peak luminosity of the IR parabola
                 model for the different classes: RQQ , SSRQ
                 (including the RG 3C 405 whose position is
                 indicated), FSRQ, and RIQ.}
         \label{hist_IR}
  \end{figure}

We define the IR luminosity $L\mathrm{(IR)}$ as the product of the luminosity
value at which each parabola peaks and the corresponding frequency ($L$(IR) =
$\nu_\mathrm{peak}\cdot L_{\nu_\mathrm{peak}}$). 
Note that this parameter does not depend on the width of the parabola. 
Only upper limits for $L\mathrm{(IR)}$ could be derived for PKS 0408$-$65
and PG 1040$-$090. The distribution of $L\mathrm{(IR)}$ is reported in
Fig.~\ref{hist_IR}. In this, and in the following histograms upper limits
are shown with arrows, one per object. The similarity in the IR luminosities
and spectra (see also Fig.~\ref{avg_SED}) in all quasars suggest a similar
origin.
  \begin{figure}
      \resizebox{8.8cm}{!}{\includegraphics{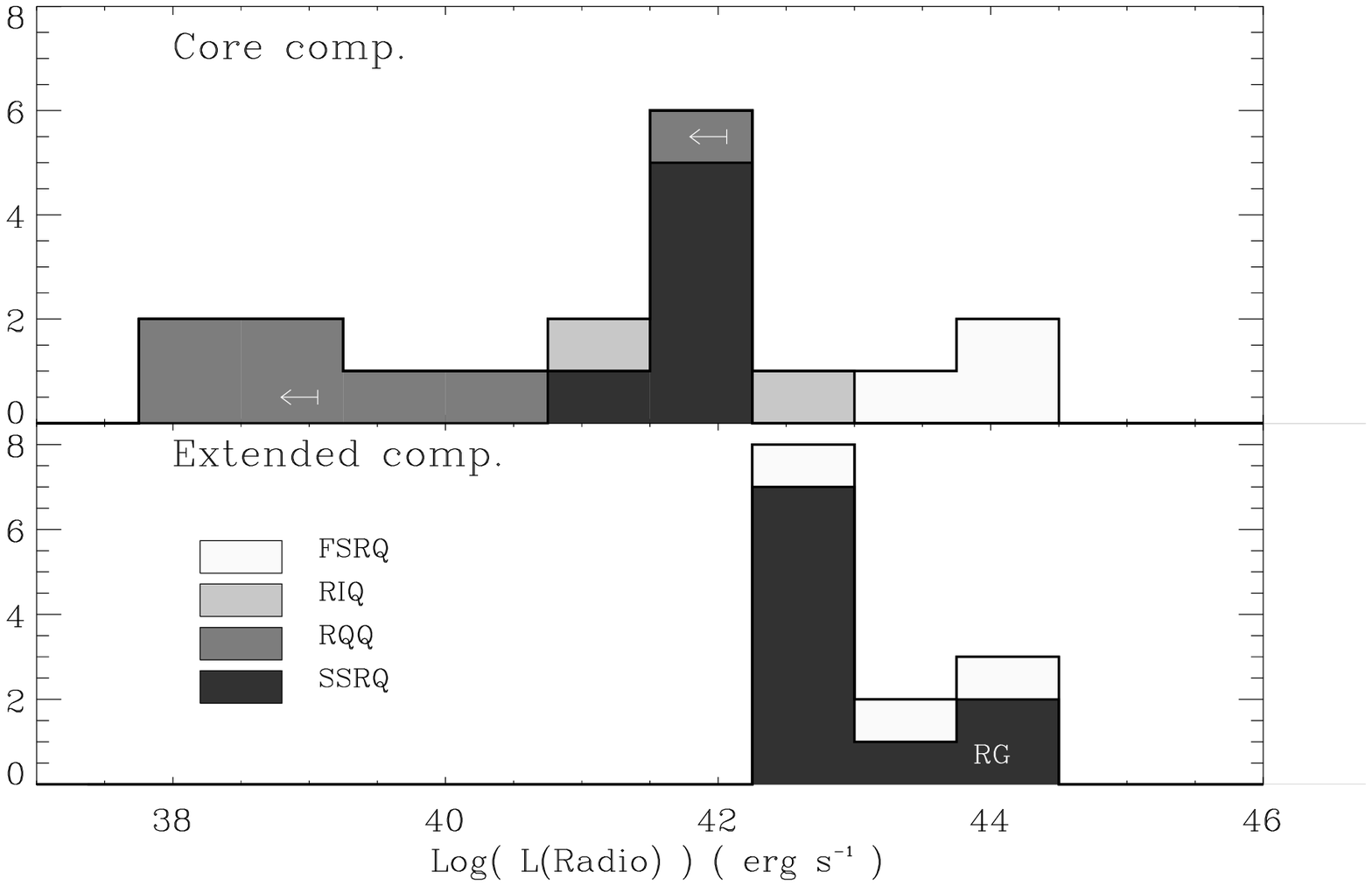}}
      \caption[]{Histogram of the core and extended components radio luminosity for the different
                 classes: RQQ, SSRQ (including the RG 3C 405
                 whose position is indicated), FSRQ, and RIQ.}
         \label{hist_jet}
  \end{figure}
  \begin{figure}
      \resizebox{8.8cm}{!}{\includegraphics{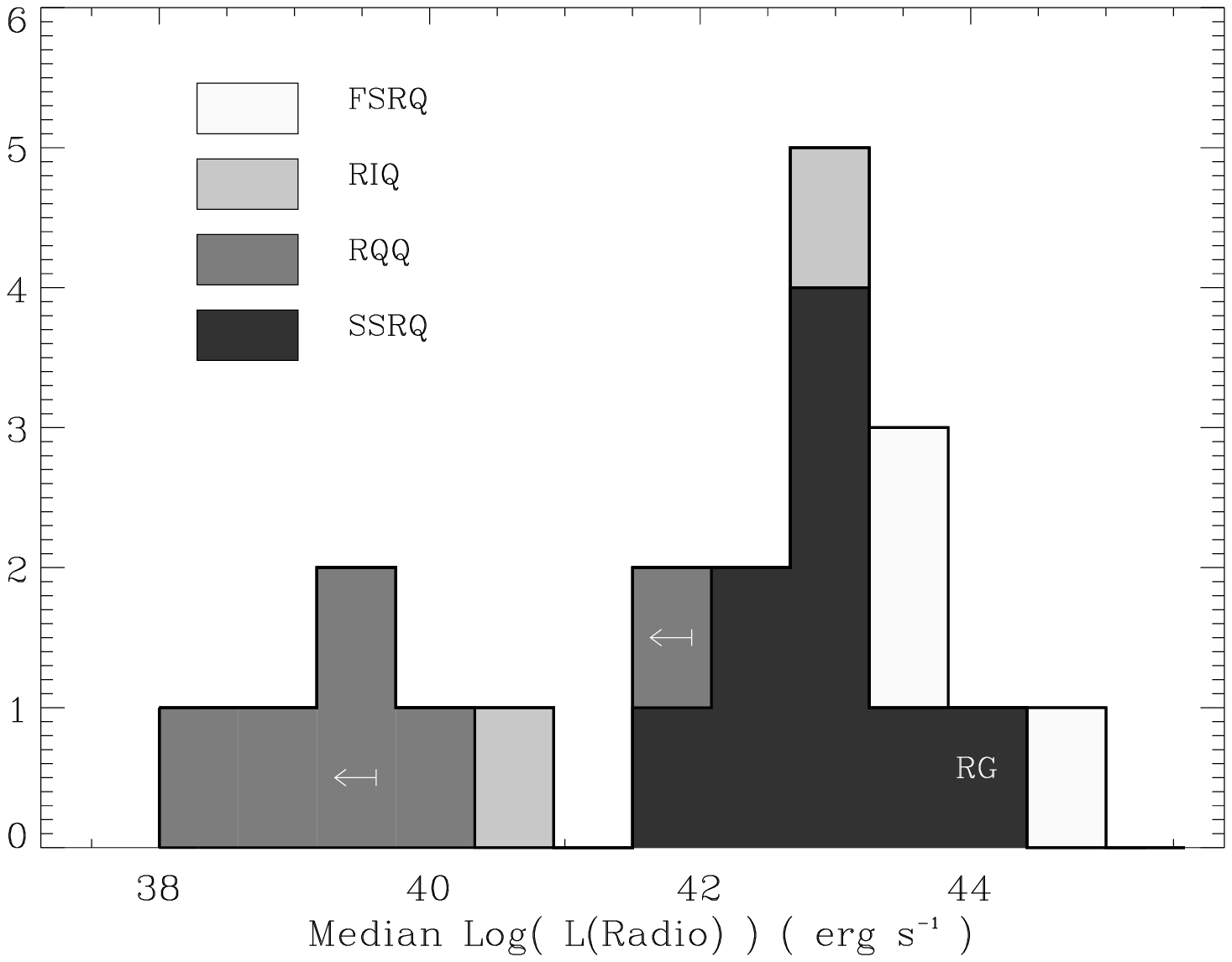}}
      \caption[]{Histogram of the median radio luminosity for the different
                 classes: RQQ, SSRQ (including the RG 3C 405
                 whose position is indicated), FSRQ, and RIQ.}
         \label{hist_rad_med}
  \end{figure}
  \begin{figure}
      \resizebox{8.8cm}{!}{\includegraphics{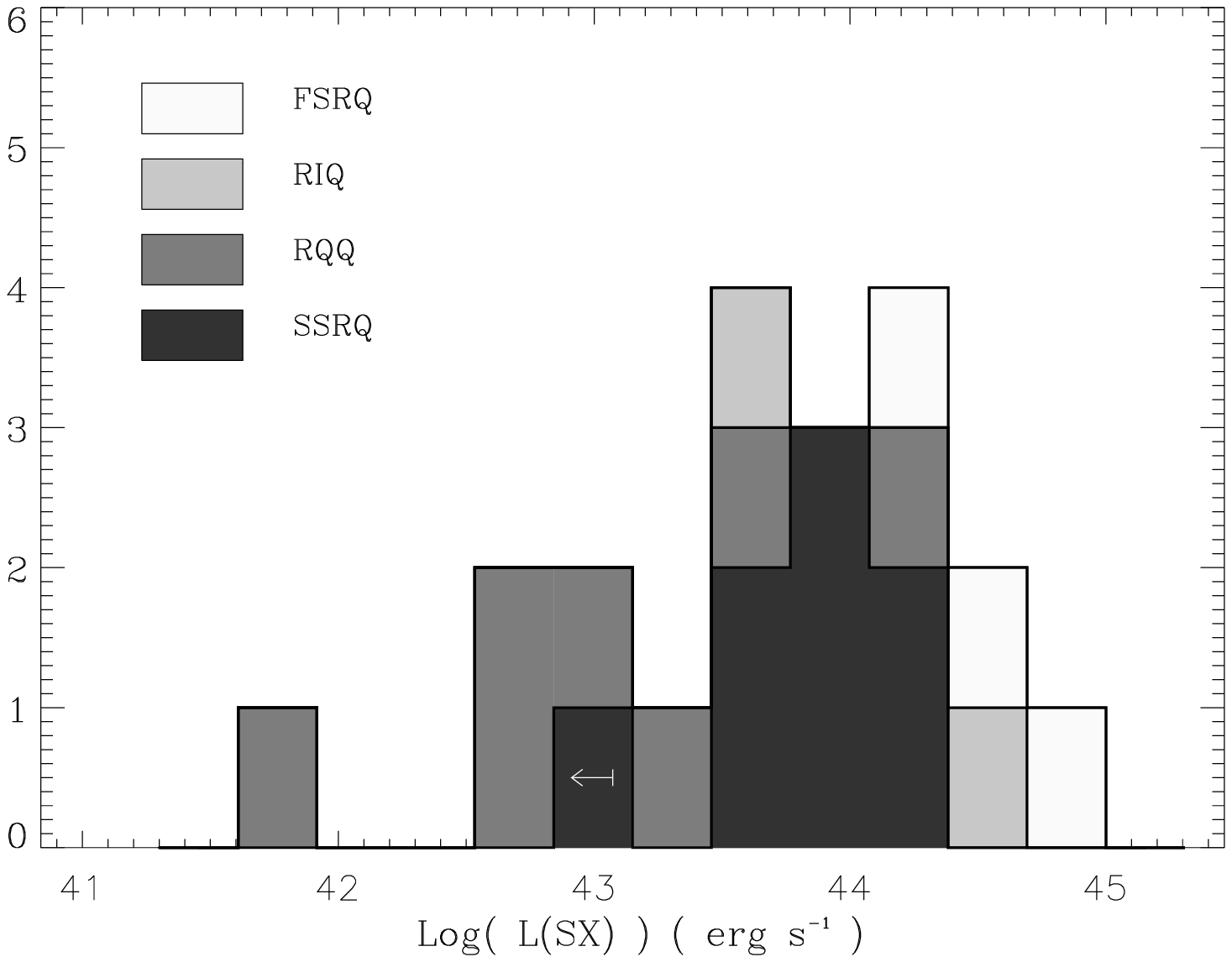}}
      \caption[]{Histogram of the soft X-ray luminosity at 1 keV for the
                 different classes: RQQ, SSRQ, FSRQ, and RIQ.}
         \label{hist_sx}
  \end{figure}

The radio emission in the RLQ arises from two very different spatial
scales, the core and extended components. We calculated the average of
$\nu$L$_{\nu}$ over the rest-frame interval 5--9 GHz for each spatial
component in all of the RLQ, except PG 1354+213 and HS 1946+7658, which were
undetected at these frequencies. Fig.~\ref{hist_jet} displays histograms for
the two components separately, and Fig.~\ref{hist_rad_med} shows the
distribution of the median of all measured $\nu$L$_{\nu}$ over the same
frequency range, without component distinction. The distribution of the
median radio luminosity is bi-modal (Fig.~\ref{hist_rad_med}). However, if
we consider only core radio luminosities (top panel of Fig.~\ref{hist_jet}),
the SSRQ radio luminosity distribution shifts towards lower values, making a
continuous distribution, rather than a bi-modal one, but without
overlapping. The contribution from the extended components are very similar
in FSRQ and SSRQ (bottom panel of Fig.~\ref{hist_jet}). In the following
analysis we will consider only the core luminosity $L\mathrm{(Radio)}$.
When the core luminosity is not available (PKS 0408$-$65, B2 1721+34, 3C
405, and PG 2308+098), we report an upper limit corresponding to the average
radio luminosity relative to the extended component.

In the soft X-ray, we define $L$(SX) as $\nu$L$_{\nu}$ with $\nu$
corresponding to 1 keV in the observer's rest-frame. The distribution of
$L$(SX) for each class is reported in Fig.~\ref{hist_sx}. In the soft X-ray,
no data are available for 3C 405, and PKS 0408$-$65, and only an upper limit
is available for PG 1004+130.

\subsection{Origin of the observed luminosities}

The main factors determining the observed luminosities are: the energy
emitted by the central engine (AGN); the amplification due to Doppler
boosting in a relativistic jet; and the contribution from a starburst.  We
will estimate the role of each of these parameters in producing the SEDs
through the comparison of the observed radio, IR, and soft X-ray
luminosities, represented by $L$(Radio), $L$(IR) and $L$(SX), respectively
(see section~\ref{multibands} for their definition).

\subsubsection{Orientation effects in RLQ}
\label{orient}

The orientation of the beamed emission can be estimated from the radio
core fraction R. This quantity, defined as the ratio between the core radio
luminosity and the luminosity of the extended radio emission at 5--9 GHz in
the rest frame, serves as an orientation indicator of the radio source with
respect to the observer, measuring the relative strength of the core
component (\cite{Hes95}). The core flux was not available for three SSRQ and
one RG, thus the parameter R was not computed. In the case of FSRQ we
computed the luminosity of the extended component in the frequency range
5--9 GHz by extrapolating the power law observed at low frequencies (power
law index $\alpha_\mathrm{0}$ in Table~\ref{nonthermal_param}).
\begin{figure}
      \resizebox{8.8cm}{!}{\includegraphics{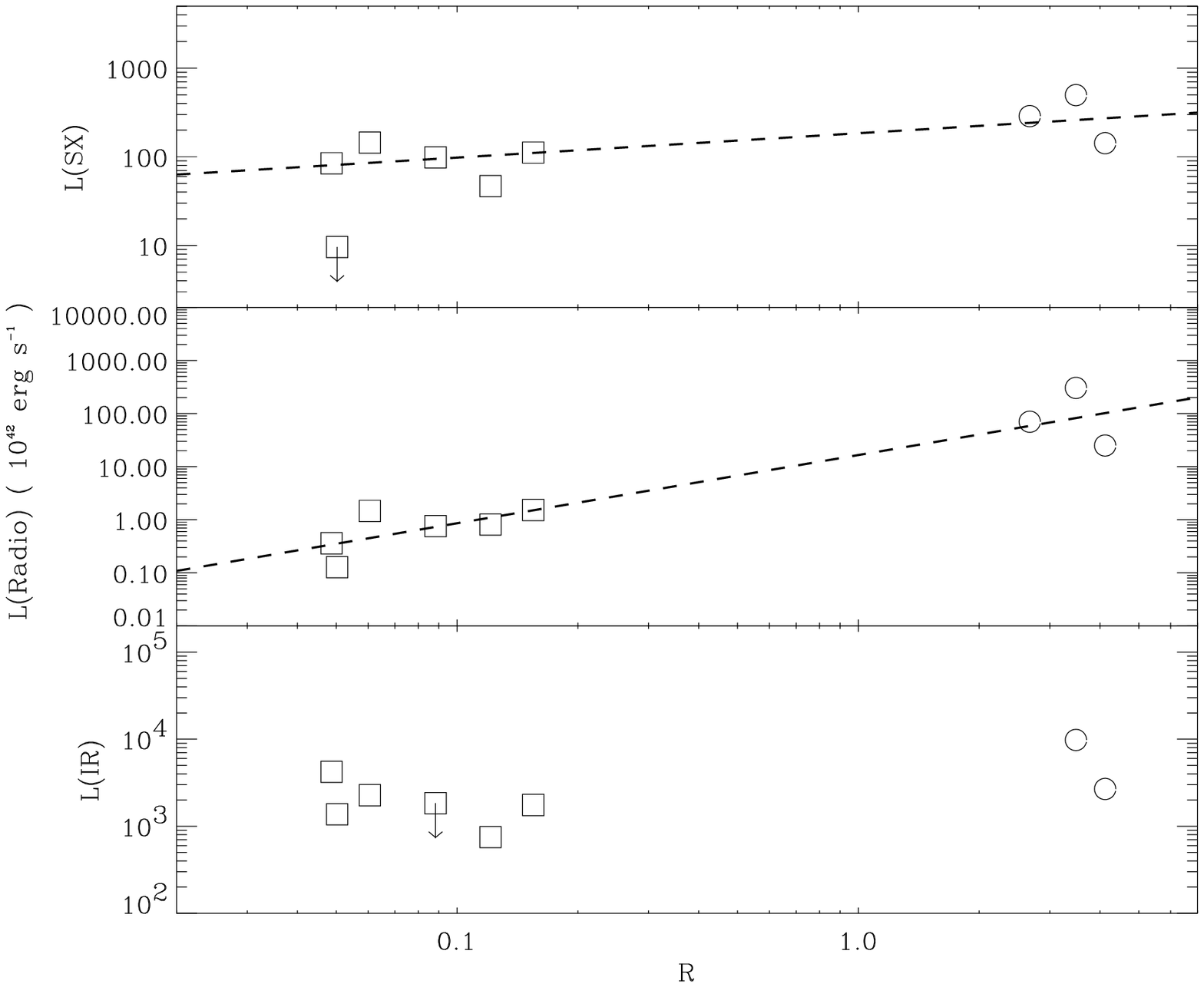}}
      \caption[]{Luminosity emitted in the soft X-ray, radio and IR domains
       compared to the parameter R given by the ratio between the radio
       luminosity of the core and of the extended components. The
       different symbols correspond to different classes of objects:
       squares for SSRQ, and circles for FSRQ. The dashed lines are the best
       fit lines.}
         \label{plot_lum_beam}
\end{figure}

The FSRQ are well separated from SSRQ in the distribution of the ratio R
(see Fig.~\ref{plot_lum_beam}). This difference permits us to estimate
the enhancement factor of the beamed emission after a few considerations.
First, the observed radio emission in RLQ is mainly produced by the jet and
its core rather than a starburst, since star-emitting ULIRG have much lower
radio luminosity than RLQ (\cite{Colina95}). Second, we assume that the
radio source is intrinsically identical in FSRQ and SSRQ, and that the
difference in their radio emission is due only to the orientation of the
beamed emission. After these approximations we can write
\begin{equation}
L\mathrm{(Radio\;Core)}_\mathrm{FSRQ} =
A \cdot L\mathrm{(Radio\;Core)}_\mathrm{SSRQ}
\label{rad_core}
\end{equation}
where $A$ is the amplification factor of the beamed emission. Since the
luminosity of the extended components are the same for the flat and steep
radio quasars (see above), using equation (\ref{rad_core}) we can derive a
relation between the parameters $R_\mathrm{FSRQ}$ and $R_\mathrm{SSRQ}$:
$R_\mathrm{FSRQ}$ = $A\cdot R_\mathrm{SSRQ}$. Replacing the observed values
of $R_\mathrm{SSRQ}$ ($\simeq$ 0.05--0.15), and $R_\mathrm{FSRQ}$ ($\simeq$
3--4) (see Fig.~\ref{plot_lum_beam}) in the above relation yields $A$
$\simeq\,$20--80.

Fig.~\ref{plot_lum_beam} displays $L$(Radio), $L$(SX), and $L$(IR) versus
the core fraction R. Linear correlation results for these relations are
reported in Table~\ref{Corr_ranks}, where the parameter pairs are reported
in the first two columns, the number of data pairs in the third, in column 4
the linear correlation rank (r$_\mathrm{l}$), and in column 5 the associated
probability to have such a correlation rank from uncorrelated values
(P$_\mathrm{l}$).
   \begin{table}
     \caption[]{Correlation test results for FSRQ and SSRQ}
     \label{Corr_ranks} 
      \begin{tabular}{@{}l l r c@{}c@{}} 
      \hline 
\noalign{\smallskip}
\multicolumn{1}{c}{X} & 
\multicolumn{1}{c}{Y} &
\multicolumn{1}{c}{N} &
\multicolumn{1}{c}{$r_\mathrm{l}$} &
\multicolumn{1}{c}{P$_\mathrm{l}$ (\%)}\\
\hline 
\noalign{\smallskip}
 Log($L$(IR))     & Log(R) &  7 &   0.53  &    25. \\
 Log($L$(Radio))  & Log(R) &  9 &   0.94  &   0.03 \\
 Log($L$(SX))     & Log(R) &  8 &   0.73  &    6.0 \\
\noalign{\smallskip}
\hline 
\end{tabular}\\
\end{table}

Higher radio and soft X-ray luminosities are observed in objects with higher
values of the radio core fraction R (when the jet points towards us). 
The orientation effect is more important in the radio domain, as shown by
the stronger correlation, than in the soft X-ray, and negligible in the IR.
This implies that the radio core and a fraction of the total emitted soft
X-ray luminosities are emitted anisotropically. We furthermore verified
that R is not correlated with the redshift and thus that the above result is
not an artifact of distance related biases in the measurement of R.

Assuming that the soft X-ray source is intrinsically identical in FSRQ and
SSRQ, the observed difference in L(SX) arises from the orientation of the
fraction, $f$, that is beamed. If the fraction of emitted radiation that is
beamed is enhanced by a factor A, identical to that of the radio emission,
the following relation between the soft X-ray luminosity in FSRQ and that in
SSRQ will be valid:
\begin{equation}
L\mathrm{(SX)}_\mathrm{FSRQ} =
A\cdot f\cdot L\mathrm{(SX)}_\mathrm{SSRQ} + (1 - f)\cdot L\mathrm{(SX)}_\mathrm{SSRQ}.
\end{equation}
Using the average value of the ratio
$L\mathrm{(SX)}_\mathrm{FSRQ}/L\mathrm{(SX)}_\mathrm{SSRQ}$ that is $\approx 3.2
\pm 2.2$, and the range of values obtained for the factor $A$, we
derive a fraction $f \approx$ 3--12\% for the beamed fraction of the soft
X-ray component.

\subsubsection{SSRQ $versus$ RQQ}

The radio and the soft X-ray luminosities are mainly produced by the
AGN component (see sections~\ref{avgsedpar} and~\ref{orient}). The
comparison between the luminosities emitted in the radio and soft X-ray
domains is then equivalent to a comparison of the AGN power in the two types of
quasars. The radio core emission of SSRQ is on average 200 times higher than
that of RQQ (see Fig.~\ref{hist_jet}) and the soft X-ray luminosity is on
average 8 times higher in SSRQ than in RQQ (see Fig.~\ref{hist_sx}). Since
the SSRQ show luminosities higher than RQQ not only at radio and soft X-ray
energies, but also in the hard X-ray domain (\cite{Lawson97}), we argue that
the bolometric AGN luminosity is much higher in SSRQ than in RQQ. The
difference in the AGN power should be observable at all frequencies where
the AGN emission dominates. We have already pointed out the similarity in IR
luminosities and spectra of SSRQ and RQQ (see Figs.~\ref{avg_SED}
and~\ref{hist_IR}). This similarity suggests that the origin of the dominant
IR component is not AGN-related. The candidate is then a starburst.

\begin{figure}
       \resizebox{8.8cm}{!}{\includegraphics{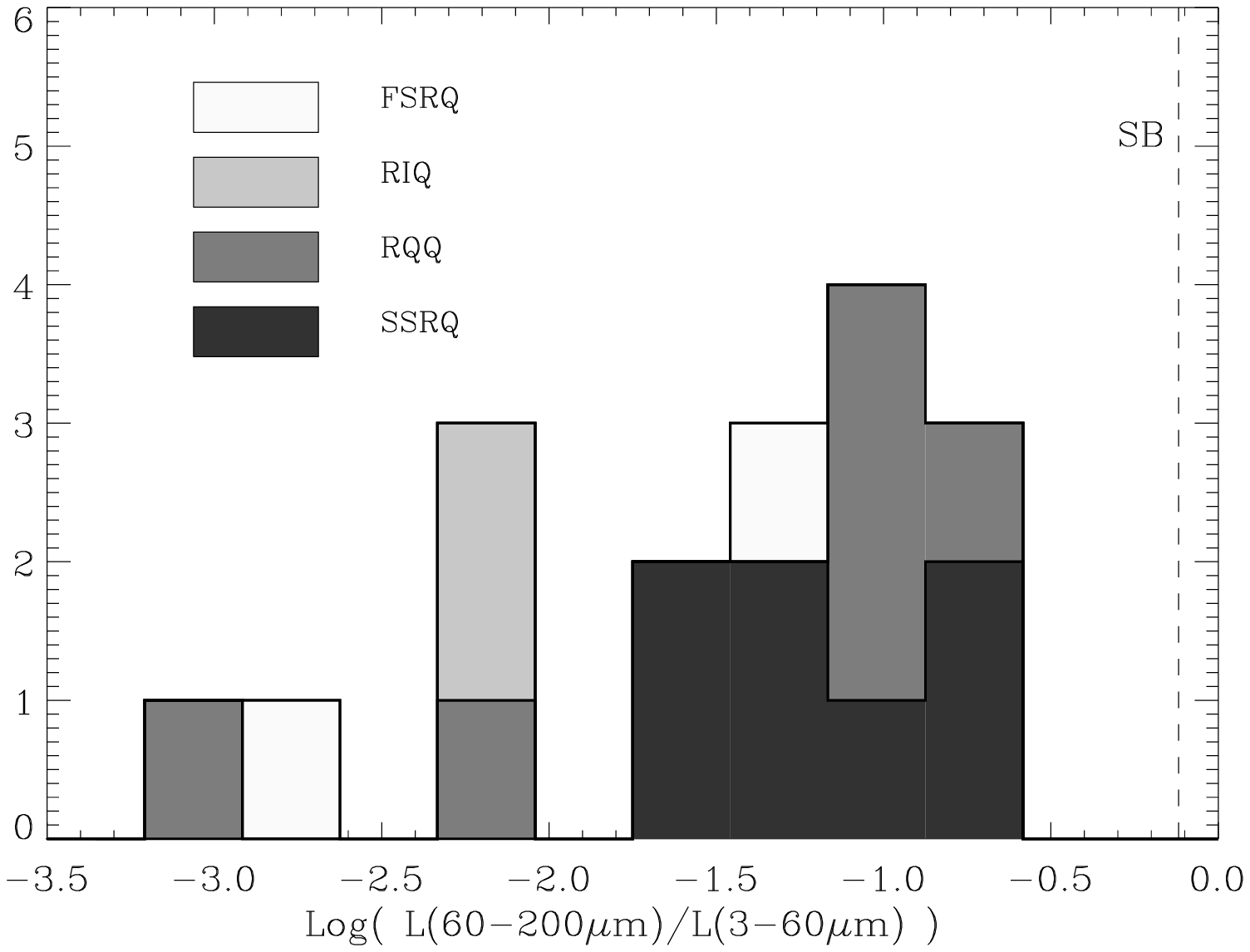}}
      \caption[]{Histogram of the ratio $L$(60-200$\mu$m)/$L$(3-60$\mu$m)
                 for the different classes: RQQ, SSRQ, FSRQ, and RIQ. The
                 dashed line represents the value derived from the average
                 SED of a sample of low reddening starburst galaxies.}
         \label{hist_SB_AGN}
  \end{figure}

Some indication of the dominant IR emission mechanism can be gleaned from
the shape of the SED.  An AGN can emit a significant fraction, often a
majority, of its infrared luminosity at shorter wavelengths, $\lambda$ $<$
60$\mu$m, as long as the obscuring columns are not so large as to be
optically thick at these wavelengths. Starburst dominated galaxies, on the
other hand, produce the bulk of their infrared emission at $\lambda$ $>$
60$\mu$m. The ratio of the luminosities in these two wavelength regimes,
$L$(60-200$\mu$m)/$L$(3-60$\mu$m), thus provides a rough estimate of the
primary driver of the infrared component. A histogram of this ratio is
presented in Fig.~\ref{hist_SB_AGN}, in which only the sources having at
least two grey body components with T $<$ 1000 K are included. For
comparison, the luminosity ratio from an average SED of low reddening
starburst galaxies (\cite{Schmitt97}) is also indicated. All of the AGN in
the sample have infrared luminosity ratios less than the starburst fiducial
value (=0.76) by a factor or four or more (the maximum ratio is 0.20
corresponding to 27\% of starburst contribution), suggesting that the
infrared in these sources is dominated by the central engine. The RQQ and
SSRQ have similar average ratios, 8\% of the total IR emission is produced
by a starburst.

Since the dominant IR source is the AGN, which differs in SSRQ and RQQ,
other factors have to be taken into account to explain the similarity of
their IR luminosities and average spectra. These factors can be the dust
covering factor, its amount and geometric distribution. A higher dust
covering factor in RQQ, due to larger total dust mass and/or a particular
geometric distribution, may account for the similar IR radiation. Aspects of
this hypothesis are attractive, since the presence of more circumnuclear
material may be related to the physical conditions that hinder the jet
formation and/or development in RQQ. However, if larger dust covering
factors were present in RQQ, a higher probability to observe a RQQ through
optically thick dust would be expected and a large fraction of RQQ with
absorption features in the soft X-ray would be observed.  Observations do
not support these predictions (see section~\ref{dataset}). Moreover, the
similarity in the observed IR luminosities and spectra requires a fine
tuning between the dust heating source, the dust amount and its geometric
distribution, which makes this explanation improbable. We propose another
scenario in which the dust properties (amount and distribution) and the
heating source are similar in SSRQ and RQQ. The dust distribution
contributing in the far-IR probably extends from the more external regions
of the AGN, and is predominantly heated by the optical and UV radiation
fields filling these external regions which are similar in both classes. At
relatively large distances from the centre, the AGN components are similar
in both types of quasars, towards the innermost regions, where the soft
X-rays are emitted and a jet is formed, SSRQ and RQQ become different. The
high energy photons escape from the centre without significant dust
absorption and provide an important probe of the central radiation source.
Observations indicate that the soft X-ray radiation is higher in
SSRQ (see Fig.~\ref{avg_SED}) in agreement with the proposed scenario.

From this analysis it is suggested that the main difference between RQQ, and
SSRQ takes place in the innermost nuclear regions where the emitted power is
higher in SSRQ than in RQQ, while the AGN external regions (dust distribution
and optical/UV source) have similar properties in both types of quasars.

Since FSRQ and SSRQ show similar properties in the IR, once the non-thermal
contribution is subtracted, this conclusion can be also extended to FSRQ.

\section{Conclusions}
\label{conc}

Continuum observations from radio to soft X-ray energies have been presented
for a sample of 22 quasars characterized by different degrees of radio
emission. The IR data were obtained with ISOPHOT and IRAC1, and the
mm data with IRAM, SEST and SCUBA. Further IR and mm data, and some radio
and soft X-ray fluxes, were drawn from the literature. The availability of a
broad band SED for several types of quasars allows us to separate the
dominant spectral components in the radio and IR energy bands, and thus to
compare their spectral properties in different types of quasars. The
spectral analysis and the comparison of luminosities emitted in the radio,
IR, and soft X-ray energy bands yields the following results:
\begin{enumerate} 
\item {\it What is the dominant mechanisms emitting at IR energies in RLQ
      and RQQ?}

      In our quasar sample the dominant mechanism emitting in the
      far/mid-IR is thermal emission from dust heated by the optical and UV
      radiation produced in the outer regions of the AGN. A starburst
      contributes to the IR emission at different levels, but always less
      than the AGN ($\leq$27\%). The presence of thermal IR emission in FSRQ
      remains rather uncertain.  Among the three FSRQ of the sample we
      cannot derive any conclusion for two of them, and for a third one (B2
      2201+31A) the observed data suggest a dominant thermal component at
      $\lambda\,\leq\,60\,\mu$m.
\item {\it Do RLQ and RQQ have the same dust properties (temperature, source
      size, mass, and luminosity)?}

      The equilibrium temperature of dust grains, the size and the mass of
      the dust distribution, and the emitted luminosity have been evaluated
      for all quasars.  The estimated sizes of the observed dust components
      lie between 0.06\,pc and 9.0\,kpc, and the temperatures
      between 43\,K and 1900\,K.
      The total luminosity observed in the IR, obtained by integrating the
      grey body components, varies over a wide range: 2--760$\cdot$10$^{11}$
      L$_\mathrm{\odot}$. The amount of emitting dust in all types of
      quasars also varies in a broad range: 6$\cdot$10$^{4}$--4$\cdot$10$^{7}$
      M$_\mathrm{\odot}$. The distribution of any of these parameters
      does not differ significantly among the different types of quasars.
\item {\it Does an interplay between the radio and the IR components exist?}

      A bright and flat non-thermal component can be sufficiently strong in
      the IR to mask the dust emission in some sources, particularly
      FSRQ. However, this does not mean that the dust emission is absent. 
      After subtracting an IR synchrotron component extrapolated from the
      radio, the residual IR emission had similar spectral shape and
      luminosity, regardless of the radio properties.
\end{enumerate}

These results are based on the analysis of a small sample, and have to be
confirmed by the study of larger samples. ISOPHOT has doubled the number of
quasars with IR detections in the sample presented here. A great deal of
additional progress on understanding the IR properties of quasars is
expected when all of the quasar data available in the ISO archive will be
fully analyzed and studied.

\begin{acknowledgements}

We are grateful to D. Carrillo, S. Paltani, and R. Walter for preparing the
observation program of ISOPHOT Chopper Observations. We thank the Time
Allocation Committee for awarding discretionary ISO observing time to this
project. Martin Haas, and P\'{e}ter \'{A}brah\'{a}m are acknowledged for
help in the ISOPHOT data reduction. R. Neri, and D. Nuernberger are
acknowledged for assistance with the IRAM interferometer data reduction. BJW
and EJH were supported in this project by NASA grant NAGW-3134. This
research has made use of NASA's Astrophysics Data System Abstract Service,
as well as the NASA/IPAC Extragalactic Database (NED) which is operated by
the Jet Propulsion Laboratory, California Institute of Technology, under
contract with the National Aeronautics and Space Administration.

\end{acknowledgements}

\end{document}